\documentclass[prb,showpacs,preprintnumbers,amsmath,amssymb]{revtex4}

\usepackage{graphicx}% Include figure files
\usepackage{dcolumn}% Align table columns on decimal point
\usepackage{bm}% bold math

\begin{document}

\title{Spin fluctuations and superconductivity
in noncentrosymmetric heavy fermion systems CeRhSi$_3$ and
CeIrSi$_3$}

\author{Y. Tada}, 
\author{N. Kawakami},
\author{S. Fujimoto}

\affiliation{Department of Physics, Kyoto University, Kyoto 606-8502,
Japan}

\newcommand{\vecc}[1]{\mbox{\boldmath $#1$}}

\begin{abstract}
We study the normal and the superconducting properties
in noncentrosymmetric heavy fermion superconductors CeRhSi$_3$ and CeIrSi$_3$.
For the normal state, we show that experimentally observed 
linear temperature dependence of the resistivity is understood
through the antiferromagnetic spin fluctuations near the 
quantum critical point (QCP) in three dimensions.
For the superconducting state, 
we derive a general formula to calculate the 
upper critical field $H_{c2}$,
with which we can treat the Pauli and the
orbital depairing effect on an equal footing.
The strong coupling effect
for general electronic structures is also taken into account.
We show that
the experimentally observed features in $H_{c2}\parallel \hat{z}$,
the huge value up to $30$(T), the downward curvatures, and
the strong pressure dependence,
are naturally understood
as an interplay of the Rashba spin-orbit interaction
due to the lack of inversion symmetry
and the spin fluctuations near the QCP.
The large anisotropy
between $H_{c2}\parallel \hat{z}$ and $H_{c2}\perp \hat{z}$ is
explained in terms of the spin-orbit interaction. 
Furthermore, a possible realization of the Fulde-Ferrell-
Larkin-Ovchinnikov state for $H\perp \hat{z}$ is studied.
We also examine effects of spin-flip scattering processes
in the pairing interaction and those of 
the applied magnetic field on the spin fluctuations.
We find that the above mentioned results are robust against these effects.
The consistency of our results strongly supports the scenario that
the superconductivity in CeRhSi$_3$ and CeIrSi$_3$ is mediated
by the spin fluctuations near the QCP.
\end{abstract}

\pacs{Valid PACS appear here}

\maketitle
%%%%%%%%%%%%%%%%%%%%%%%%%%%%%%%%%%%%%%%%%%%%%%%%%%%%%%%%%%%%%%%%%
\section{Introduction}
\label{sec:intro}
%%%%%%%%%%%%%%%%%%%%%%%%%%%%%%%%%%%%%%%%%%%%%%%%%%%%%%%%%%%%%%%%
In noncentrosymmetric heavy fermion superconductors,
in addition to strong electron correlation, 
there exists another key property,
the anisotropic spin-orbit (SO) interaction due to the 
lack of inversion symmetry.
The anisotropic SO interaction plays important roles
both in the normal and the superconducting state, and is expected
to lead to many interesting phenomena.~\cite{pap:Edelstein89,
pap:Edelstein95,pap:Yip,pap:Gorkov,pap:Frigeri,pap:Samokhin04,
pap:Samokhin05,pap:Mineev05,pap:Mineev07}
For such phenomena,
electron correlation is quite important,
because it can largely enhance the effect of the SO interaction.
The interplay of the anisotropic SO interaction
and electron correlation is truly an unique nature
in noncentrosymmetric heavy fermion compounds.~\cite{pap:FujimotoPRB,
pap:FujimotoJPSJ1,pap:FujimotoJPSJ2}
In particular, such an interplay in the superconducting
state has been attracting particular interest.
In this context, especially,
CeRhSi$_3$~\cite{pap:Kimura,pap:Muro,pap:KimuraJPSJ} 
and CeIrSi$_3$~\cite{pap:Sugitani,
pap:Okuda} are promising candidates for the interplay,
because they are considered to be located near the antiferromagnetic (AF)
QCPs around which strong correlations 
through the spin fluctuations are essential.

CeRhSi$_3$ and CeIrSi$_3$ are AF metals at ambient pressure,
and begin to exhibit superconductivity at some 
critical pressures $P_c$ where
the N\'eel temperatures seem to be suppressed to absolute zero.
According to the neutron experiments for CeRhSi$_3$, the AF ordering vector is
$\vecc{Q}=(\pm 0.43\pi,0,0.5\pi),
(0,\pm 0.43\pi,0.5\pi)$ and the nature of the AF order is 
spin density wave-like.~\cite{pap:Aso}
This is different from CePt$_3$Si in which 
the AF seems to have localized
nature and the superconductivity coexists with it even at zero applied
pressure.~\cite{pap:Bauer}
In NMR experiments in CeIrSi$_3$, $1/T_1\propto T/(T+\theta)^{1/2}$ is 
observed near the critical pressure, which is a characteristic
behavior of the systems with 3-dimensional (3D) AF spin 
fluctuations.~\cite{pap:Mukuda,pap:SCR1,pap:SCR2}
In addition, the resistivity $\rho$ in both CeRhSi$_3$ and 
CeIrSi$_3$ above the superconducting
transition temperatures $T_c$ in some pressure regions near the QCP
shows the anomalous $T$-linear dependence which is different from
$\rho \sim T^2$ in canonical Fermi liquids.~\cite{pap:KimuraJPSJ,
pap:Sugitani}

The QCP related phenomena are observed also in the superconducting
state.
The large jump in the heat capacity at $T_c$ in CeIrSi$_3$
can be attributed to the strong coupling effect
due to the spin fluctuations.~\cite{pap:Tateiwa}
It has strong pressure dependence and is largely enhanced near $P_c$.
The most striking phenomena which would be related to the quantum criticality
in CeRhSi$_3$ and CeIrSi$_3$ appear in the behaviors of 
the upper critical fields $H_{c2}$ when the applied magnetic field is 
parallel to $z$-axis.~\cite{pap:Kimura_Hc2,pap:Settai}
The remarkable features of the experimental results are as 
follows.
(i) As the pressure approaches a critical value,
$H_{c2}$ exhibits extremely high value which exceeds the orbital
limit as well as the Pauli limit estimated by the conventional
BCS theory. The observed 
$H_{c2}\sim 30$(T) is the highest value among the heavy fermion
superconductors ever discovered, although $T_c$ is merely
$T_c\sim 1$(K).
(ii) $H_{c2}$ curves have downward curvatures and
the increase is accelerated as the temperature is decreased,
making a sharp contrast to any other superconductors
in which the increase of $H_{c2}$ becomes slower as $T$ is
decreased.
(iii) $H_{c2}$ increases very rapidly as the pressure approaches
the critical value, while the pressure dependence of $T_c$
is moderate.
These characteristic features strongly suggest that there exists
a deep connection between the 
superconductivity and the magnetic quantum criticality.
In the previous study, the present authors 
have shown that these experimental
results are well explained as an interplay of the Rashba SO interaction
due to the lack of inversion symmetry and the spin fluctuations
near the QCP.~\cite{pap:TadaPRL}

On the other hand, $H_{c2}$ for in-plane fields 
differs from $H_{c2}\parallel \hat{z}$ in some important features.
$H_{c2}\perp \hat{z}$ is merely less than $10$(T) and 
its pressure dependence is moderate, and
the $H_{c2}$ curves exhibit usual upward curvatures.~\cite{pap:Kimura_Hc2,
pap:Settai}
This anisotropy in $H_{c2}$ would be related to the Rashba SO interaction,
since the Fermi surface is asymmetrically distorted by the in-plane field
and the Pauli depairing effect plays essential roles.
By contrast, 
the renormalization of the quasiparticle velocity by the 
spin fluctuations is almost
isotropic, resulting in the enhanced orbital limiting field
in all directions of the applied field.
Another interesting phenomenon in the noncentrosymmetric
superconductors in applied magnetic fields
is the helical vortex phase which has been
discussed theoretically. ~\cite{pap:Kaur,pap:Yanase,
pap:Dimitrova,pap:Agterberg,pap:Samokhin,pap:Matsunaga,pap:Hiasa,pap:Mineev}
In a helical vortex phase, the superconducting gap function
is modulated in real space, $\Delta (\vecc{R})\sim \exp 
(i\vecc{Q}\cdot \vecc{R})\Delta$ with the modulation vector
$Q\sim (\alpha/\varepsilon_F) \mu_BH/v_F$, where $\alpha,v_F,\varepsilon_F$ 
and $\mu_B$ are the 
strength of the SO interaction, the Fermi velocity,
the Fermi energy, and the Bohr magneton, respectively.
For 3D Rashba superconductors for $H\perp \hat{z}$, 
however, it is pointed out that this phase modulation is just
a translational shift of the vortex lattice and has no
physical importance.~\cite{pap:Matsunaga,pap:Hiasa,pap:Mineev}
Several authors also have discussed a spatially modulated
superconducting state under magnetic fields with a large 
$Q\sim \mu_BH/v_F$
which is continuously connected from 
$Q\sim (\alpha/\varepsilon_F) \mu_BH/v_F$.~\cite{pap:Yanase,
pap:Dimitrova,pap:Samokhin,pap:Hiasa}
This large $Q$ state corresponds to the Fulde-Ferrell-Ovchinnikov-Larkin
(FFLO) state.~\cite{pap:FF,pap:LO}
The stability of the modulating superconducting state with the large $Q$
depends on the relative strength of the orbital depairing effect to
the Pauli depairing effect in the compounds.

In this paper, we study
the normal and the superconducting properties
in noncentrosymmetric superconductors CeRhSi$_3$ and
CeIrSi$_3$.
We examine the anomalous $T$-linear dependence of
the resistivity in the normal state.
In the previous studies,\cite{pap:SCR1,pap:SCR2} at very low temperatures,
$\rho \sim T^{3/2}$ is predicted for 3D AF spin fluctuations.
The temperature dependence of the resistivity for 3D AF spin
fluctuations have been studied in detail by several 
authors~\cite{pap:Rosch,pap:RoschPRB,pap:Onari}
We, here, show that $\rho \sim T$ in CeRhSi$_3$ and CeIrSi$_3$
is actually due to the 3D AF spin fluctuations.
For the superconducting state, the upper critical fields both for
$H\parallel \hat{z}$ and $H\perp \hat{z}$ are investigated.
For the calculation of $H_{c2}$, 
we derive a general formula which enables us to treat the Pauli
and the orbital depairing effects on an equal footing.
We can also take into account the strong coupling effect
for a given electronic structure.
We calculate $H_{c2}$ on the basis of the scenario that
the superconductivity in CeRhSi$_3$
and CeIrSi$_3$ is mediated by the spin fluctuations, and
show that the experimental features are well explained
as an interplay of the spin fluctuations and the Rashba SO
interaction.
Although the formula is applicable for general models,
we use a phenomenological model to calculate $H_{c2}$ and
neglect the following two points in the model.
One is the scattering processes in the pairing interaction
in which spins of quasiparticles are flipped by the Rashba
SO interaction.
It is pointed out that such processes can enhance the 
admixture of the singlet and the triplet 
superconductivity,~\cite{pap:Yanase08,pap:Takimoto}
and the strength of the admixture affects 
$H_{c2}\perp \hat{z}$.~\cite{pap:Hiasa}
We show that the admixture is still small in CeRhSi$_3$
and CeIrSi$_3$ even if we
include the spin-flip scattering processes.
The other point is the applied field dependence of the
spin fluctuations.
Because the applied field is so large in CeRhSi$_3$ and CeIrSi$_3$
especially for $H\parallel \hat{z}$ that one might think that
the spin fluctuations are suppressed and they cannot contribute to
the enhancement of $H_{c2}$.
We show that the spin fluctuations are robust against the 
applied field $H$ up to the strength of the Rashba SO interaction,
$\mu_BH\lesssim \alpha$, because the Rashba SO interaction tends to
fix the directions of spins on the Fermi surface and it
competes with the Zeeman effect.
The consistency of our results with the experiments strongly supports
the scenario that the superconductivity in CeRhSi$_3$ and 
CeIrSi$_3$ is mediated by the spin fluctuations near the AF QCP.

This paper is organized as follows.
In Sec.\ref{sec:normal}, we study the experimentally 
observed $T$-linear dependence of the resistivity.
In Sec.\ref{sec:Eliashberg}, a general formula for the
calculation of $H_{c2}$ is derived from the Eliashberg 
equation.
The characteristic features of $H_{c2}$ in 
CeRhSi$_3$ and CeIrSi$_3$ are well explained with the 
use of the formula in Sec.\ref{sec:Hc2}.
We discuss, in Sec.\ref{sec:2points}, the spin-flip
scattering processes in the pairing interaction and the 
magnetic field dependence of the spin fluctuations which
are not included in the approximation
used for the computation of $H_{c2}$.
The summary is given in
Sec.\ref{sec:summary}.

%%%%%%%%%%%%%%%%%%%%%%%%%%%%%%%%%%%%%%%%%%%%%%%%%%%%%
\section{resistivity in normal state}
\label{sec:normal}
%%%%%%%%%%%%%%%%%%%%%%%%%%%%%%%%%%%%%%%%%%%%%%%%%%%%%
In this section, we discuss the temperature dependence of
the resistivity
near the AF QCP in CeRhSi$_3$ and CeIrSi$_3$.
In CeIrSi$_3$, NMR $1/T_1$ behaves as $1/T_1\propto
T/\sqrt{T+\theta}$
in some pressure regions, which means that the character of the
spin fluctuations is 3D 
antiferromagnetic.~\cite{pap:Mukuda,pap:SCR1,pap:SCR2}
In noncentrosymmetric systems, however, spin fluctuations are
not isotropic due to the anisotropic spin-orbit interaction.
The anisotropy in
the noninteracting susceptibility $\hat{\chi}^0$ is
of the order of $\alpha/\varepsilon_F\ll 1$, where
$\alpha$ is the strength of the spin-orbit interaction and
$\varepsilon_F$ is the Fermi energy.
Actually, in Sec.\ref{subsec:susc}, we show that the anisotropy among 
$\chi_{xx},\chi_{yy}$ and $\chi_{zz}$ is very small
within the random phase approximation.
Therefore, we can neglect the anisotropy of the spin fluctuations
for the discussion of the resistivity and
$H_{c2}$ in CeRhSi$_3$ and CeIrSi$_3$.

For the systems with 3D AF spin fluctuations,
the resistivities are expected to be $\rho \sim T^{3/2}$
according to the previous studies.~\cite{pap:SCR1,pap:SCR2}
In CeRhSi$_3$ and CeIrSi$_3$, however, the temperature
dependence is $\rho \sim T$ near the AF QCP.~\cite{pap:KimuraJPSJ,
pap:Sugitani}
The resistivity due to the 3D AF spin fluctuations was
discussed by several authors, 
and $\rho \sim T$ behavior was found for the weakly 
disordered systems~\cite{pap:Rosch,pap:RoschPRB}
and the clean systems.~\cite{pap:Onari}
Here, we show that the $T$-linear resistivity in CeRhSi$_3$ and
CeIrSi$_3$ is naturally understood in terms of
the 3D AF spin fluctuations, and this behavior has basically
nothing to do with the lack of inversion symmetry.

CeRhSi$_3$ and CeIrSi$_3$ are heavy fermion systems 
with Kondo temperature $T_{\rm K}\sim $50-100(K) which is much
higher than the superconducting
transition temperature 
$T_c\sim 1$(K).~\cite{pap:Kimura,pap:Sugitani,pap:Muro98}
Therefore, to study the properties at $T=$1-10(K),
we consider the low energy quasiparticles mainly 
formed by $f$-electrons through the hybridizations with the
conduction electrons.
We use the following single band model for the low energy quasiparticles
with the asymmetric spin-orbit interaction
%%%%%%%%%%%%%%%%%%%%%%%%%%%%%%%%%%%
\begin{eqnarray}
S&=&S_0+S_{\rm SF},\label{eq:action}\\
S_0&=&
\sum_kc^{\dagger}_k[-i\omega_n+\varepsilon_0(\vecc{k})]c_k
+\sum_kc^{\dagger}_k
\alpha \vecc{\mathcal L}_0(\vecc{k},\vecc{H})
\cdot \vecc{\sigma}c_k,\\
S_{\rm SF}&=&-\sum_{kk^{\prime}q} \frac{g^2}{6}\chi(q)
\vecc{\sigma}_{\alpha \alpha^{\prime}}
\cdot \vecc{\sigma}_{\beta \beta^{\prime}}
c^{\dagger}_{k+q\alpha}c_{k\alpha^{\prime}}
c^{\dagger}_{k^{\prime}-q\beta}c_{k^{\prime}\beta^{\prime}},
\label{eq:action_SF}
\end{eqnarray}
%%%%%%%%%%%%%%%%%%%%%%%%%%%%%%%%%%%
where $c_{k\sigma}^{(\dagger)}$ is the annihilation (creation) 
operator of the Kramers doublet of the $\Gamma_7$ state.
$S_{\rm SF}$ is introduced phenomenologically and 
represents the interaction between the quasiparticles 
by the strong spin fluctuations near the AF QCP.
Since CeRhSi$_3$ and CeIrSi$_3$ have body-centered 
tetragonal lattice structures with lattice spacing 
$1:1:2$,~\cite{pap:Muro,pap:Okuda}
the dispersion relation $\varepsilon_0(\vecc{k})$ and
the Rashba type SO interaction 
are approximated by  
%%%%%%%%%%%%%%%%%%%%%%%%%%%%%%%%%%%%%%
\begin{eqnarray}
\varepsilon_0(\vecc{k})&=&-2t_1(\cos k_xa+\cos k_ya)+4t_2\cos k_xa\cos k_ya
\nonumber \\
&&-8t_3\cos(k_xa/2) \cos(k_ya/2) \cos k_za-2t_4\cos 2k_za-\mu, 
\label{eq:dispersion}\\
\vecc{\mathcal L}_0(\vecc{k},\vecc{H})
&=&(\sin k_ya,-\sin k_xa-\mu_BH_y/\alpha,-\mu_BH_z/\alpha),
\label{eq:Rashba}
\end{eqnarray}
%%%%%%%%%%%%%%%%%%%%%%%%%%%%%%%%%%%%%%
where $a$ is the lattice constant
and $\mu$ is the chemical potential.
Although $\vecc{H}=0$ in this section, we include the Zeeman
effect in the action for the later discussion.
We fix the parameters as
$(t_1,t_2,t_3,t_4,n,\alpha)=(1.0,0.475,0.3,0.0,1.0,0.5)$ by taking $t_1$ as
the energy unit.
The Fermi surface determined by these parameters is 
in qualitative agreement with the band calculation
and 
can reproduce the peak structures of the momentum-dependent 
susceptibility 
observed by the neutron scattering 
experiments~\cite{pap:Aso,pap:TadaJPSJ,pap:Harima}.
Since we consider $f$-electron systems, we assume that 
the above parameters include effects of the mass renormalization
due to local spin correlations with typical energy scale 
$T_{\rm K}\sim$ 50-100 (K),
the Kondo temperature; i.e. 
$t_1\sim$ 50-100 (K).

The interactions are phenomenologically introduced through
the renormalized susceptibility $\chi(q)$,~\cite{pap:SCR1,pap:SCR2,
pap:TadaPRL,pap:Monthoux92,pap:Monthoux99}
%%%%%%%%%%%%%%%%%%%%%%%%%%%%%%%%%%%%%
\begin{eqnarray}
\chi(i\nu_n,\vecc{q})&=&\sum_{a}
\frac{\chi_0\xi^2q_0^2}{1+\xi^2(\vecc{q}-\vecc{Q}_a)^2
+|\nu_n|/(\Gamma_0\xi^{-2}q_0^{-2})},
\label{eq:chi}\\
\xi(T,\theta)&=&\tilde{\xi}\sqrt{\frac{t_1}{T+\theta}},
\end{eqnarray}
%%%%%%%%%%%%%%%%%%%%%%%%%%%%%%%%%%%%%
where $\chi_0$, $q_0$ and $\Gamma_0$ are respectively the susceptibility, 
the length scale and the energy scale of spin fluctuations 
without strong correlations.
These quantities are renormalized through the coherence length $\xi(T)$ as 
the system approaches the QCP. 
The critical exponent of $\xi$ is the mean field value $1/2$
and $\theta$ is
considered to decrease monotonically as the applied pressure
approaches the critical value for the AF 
order.~\cite{pap:SCR1,pap:SCR2}
The temperature dependence of $\xi$ 
is also consistent with the recent NMR experiment
for CeIrSi$_3$.~\cite{pap:Mukuda}
The ordering vectors are $\vecc{Q}_1=(\pm 0.43\pi,0,0.5\pi)/a,
\vecc{Q}_2=(0,\pm 0.43\pi,0.5\pi)/a$ 
according to the neutron scattering experiments for CeRhSi$_3$.~\cite{pap:Aso}
In this study, we fix the parameters in $\chi(i\nu_n,\vecc{q})$ 
as $q_0=a^{-1}$, $\Gamma_0=3.6t_1$
and $\tilde{\xi}=0.45a$. The value $\Gamma_0=3.6t_1$
is of the same order as
the Fermi energy without the effect of the spin fluctuations. 
$\tilde{\xi}$
is determined so that the maximum of $\xi$ would be $\xi_{\rm max}\lesssim
10a$ at the lowest temperature in this study, which is a reasonable
value for the AF spin fluctuations.
We note that
the coupling constant $g$ should be regarded
as an effective one renormalized
by the vertex corrections.~\cite{pap:Yonemitsu,pap:Monthoux_ver}

The Green's function is,~\cite{pap:FujimotoPRB,pap:FujimotoJPSJ1,
pap:FujimotoJPSJ2}
%%%%%%%%%%%%%%%%%%%%%%%%%%%%%%%
\begin{eqnarray}
G_{\alpha \beta}(k)&=&\sum_{\tau=\pm 1}l_{\tau \alpha \beta}(k)
 G_{\tau}(k),\\
l_{\tau \alpha \beta}(k)&=&\frac{1}{2}
\left(1+\tau \hat{\vecc{\mathcal L}}(k)
\cdot \vecc{\sigma} \right)_{\alpha \beta},\\
G_{\tau}(k)&=&\frac{1}{i\omega_n-\varepsilon_{\tau}(k)
-\Sigma_0(k)}
\end{eqnarray}
%%%%%%%%%%%%%%%%%%%%%%%%%%%%%%%%%%%%
where $\varepsilon_{\tau}(k)=\varepsilon_0(\vecc{k})+
\tau \alpha \|\vecc{\mathcal L}(k)\|,
\vecc{\mathcal L}(k)=\vecc{\mathcal L}_0(\vecc{k})+\vecc{\Sigma}(k)/\alpha,
\vecc{\hat{\mathcal L}}(k)=\vecc{\mathcal L}(k)/
\|\vecc{\mathcal L}(k)\|$
and $\|\vecc{\mathcal L}(k)\|=
\sqrt{\sum_{i=1}^3[{\mathcal L}_{i}(k)]^2}$.
Selfenergy is introduced as 
$
\Sigma_0=(\Sigma_{\uparrow \uparrow}+\Sigma_{\downarrow \downarrow})/2,
\Sigma_x=(\Sigma_{\downarrow \uparrow}+\Sigma_{\uparrow \downarrow})/2,
\Sigma_y=(\Sigma_{\downarrow \uparrow}-\Sigma_{\uparrow \downarrow})/2i,
$ and
$
\Sigma_z=(\Sigma_{\uparrow \uparrow}-\Sigma_{\downarrow \downarrow})/2.
$
Up to the first order in $g^2\chi_0$,
$\Sigma_0$ and $\vecc{\Sigma}$ are expressed as
%%%%%%%%%%%%%%%%%%%%%%%%%%%%%%%%%%%%%
\begin{eqnarray}
\Sigma_0(k)&=&\frac{T}{2N}\sum_{k^{\prime}}g^2\chi(k-k^{\prime})
[G^0_{\uparrow \uparrow}(k^{\prime})+G^0_{\downarrow \downarrow}(k^{\prime})]
\label{eq:Sigma_0},\\
\Sigma_x(k)&=&\frac{T}{2N}\sum_{k^{\prime}}\frac{g^2}{3}\chi(k-k^{\prime})
[-G^0_{\downarrow \uparrow}(k^{\prime})
-G^0_{\uparrow \downarrow}(k^{\prime})],
\label{eq:Sigma_x}\\
\Sigma_y(k)&=&\frac{T}{2iN}\sum_{k^{\prime}}\frac{g^2}{3}\chi(k-k^{\prime})
[-G^0_{\downarrow \uparrow}(k^{\prime})
+G^0_{\uparrow \downarrow}(k^{\prime})],
\label{eq:Sigma_y}\\
\Sigma_z(k)&=&\frac{T}{2N}\sum_{k^{\prime}}\frac{g^2}{3}\chi(k-k^{\prime})
[-G^0_{\uparrow \uparrow}(k^{\prime})
+G^0_{\downarrow \downarrow}(k^{\prime})]
\label{eq:Sigma_z}
\end{eqnarray}
%%%%%%%%%%%%%%%%%%%%%%%%%%%%%%%%%%%%%%
where $G^0_{\alpha \beta}$ is the non-interacting Green's function.
We have neglected the constant terms in $\Sigma_{\mu}$.
In the above expression of $\Sigma_{\mu}$, the most dominant term
is $\Sigma_0$, and $\Sigma_{x,y}$ is smaller than $\Sigma_0$
by the factor $\alpha/\varepsilon_F\ll 1$, where $\varepsilon_F$
is the Fermi energy.
For Rashba superconductors, $\Sigma_z=0$ without magnetic
field.

The conductivity is calculated from the Kubo formula
%%%%%%%%%%%%%%%%%%%%%%%%%%%%%%%%%%%%
\begin{eqnarray}
\sigma_{\mu \nu}&=&\lim_{\omega \rightarrow 0}\frac{1}{\omega}
{\rm Im}K_{\mu \nu}^R(\omega),\\
K_{\mu \nu}(i\omega_n)&=&\int_0^{\beta}d\tau e^{i\omega_n\tau}
\langle TJ_{\mu}(\tau)J_{\nu}(0)\rangle,
\label{eq:sigma_munu}
\end{eqnarray}
%%%%%%%%%%%%%%%%%%%%%%%%%%%%%%%%%%%
where the current operator $J_{\mu}$ is defined as
%%%%%%%%%%%%%%%%%%%%%%%%%%%%%%%%%%%%
\begin{eqnarray}
J_{\mu}&=&e\sum_kc_k^{\dagger}\hat{v}_{k\mu}c_k,\\
\hat{v}_{k\mu}&=&\nabla_{\mu}
[\varepsilon_0(\vecc{k})+\alpha \vecc{\mathcal L}_0(\vecc{k})
\cdot \vecc{\sigma}].
\end{eqnarray}
%%%%%%%%%%%%%%%%%%%%%%%%%%%%%%%%%%%
After the analytic continuation, 
$G^A_{\tau}G^R_{\tau^{\prime}}$ has the dominant
contribution to the conductivity.
Among the four terms $\{G^A_{\tau}G^R_{\tau^{\prime}}
\}_{\tau,\tau^{\prime}=\pm 1}$,
for sufficiently large $\alpha$, the terms 
$G^A_{+}G^R_{-}$ and $G^A_{-}G^R_{+}$
have no singularity with respect to the quasiparticle damping rate.
Therefore, we can neglect them, and the resulting expression
for $\sigma_{xx}$ is
%%%%%%%%%%%%%%%%%%%%%%%%%%%%%%%%%%%%%
\begin{eqnarray}
\sigma_{xx}&=&e^2\sum_{k,\tau} {\rm tr}[\hat{v}_{kx}\hat{l}^A_{k\tau}
\hat{v}_{kx}\hat{l}^{R}_{k\tau}]
\left( -\frac{\partial f}{\partial \varepsilon}
(\varepsilon_{k\tau})\right)
\frac{1}{2\gamma_{k\tau}}
\label{eq:sigma_xx},\\
\gamma_{k\tau}&=&-[{\rm Im}\Sigma_0^{\rm R}(0,\vecc{k})+\tau {\rm Re}
\hat{\vecc{\mathcal L}}^R(0,\vecc{k})\cdot{\rm Im}\vecc{\Sigma}^R(0,\vecc{k})]
\label{eq:gamma}
\end{eqnarray}
%%%%%%%%%%%%%%%%%%%%%%%%%%%%%%%%%%%%%%
where ${\rm Re}\hat{\vecc{\mathcal L}}^R={\rm Re}\vecc{\mathcal L}^R/
\|{\rm Re}\vecc{\mathcal L}^R\|$, $l^{(R,A)}_{k\tau}=
l^{(R,A)}_{\tau}(\varepsilon_{k\tau},\vecc{k})$ and
$\varepsilon_{k\tau}=\varepsilon_{\tau}(\vecc{k})+{\rm Re}
\Sigma_0(0,\vecc{k})$.
Here we have neglected the vertex corrections which are necessary for
the current conservation law.~\cite{pap:Eliashberg,
pap:Yamada,pap:Kontani,book:Yamada}
This is because,
for the resistivity, the scatterings by the AF spin fluctuations have large
momentum transfers, and therefore, the back-flow included in the
vertex corrections does not affect the temperature dependence
of the resistivity.~\cite{book:Yamada}
As mentioned above, $\vecc{\Sigma}$ is much smaller than
$\Sigma_0$ in amplitude, and from Eqs.(\ref{eq:Sigma_0})
$\sim$(\ref{eq:Sigma_z}), the temperature dependence
of $\Sigma_0$ and that of $\vecc{\Sigma}$ are the same.
Therefore, hereafter, we neglect $\vecc{\Sigma}$ and 
take into account only $\Sigma_0$ in this study.
The resistivity in noncentrosymmetric systems
is almost the same as that in usual centrosymmetric systems.
This is different from the situations for the anomalous Hall effect,
the magnetoelectric effect and so on for which 
the Rashba SO interaction plays important roles.~\cite{pap:Edelstein95,
pap:Yip,pap:FujimotoPRB,pap:FujimotoJPSJ1,pap:FujimotoJPSJ2}

Before moving to the numerical calculation,
we show a simple analytical result for ${\rm Im}\Sigma_0$ which 
determines the qualitative behavior of $\sigma_{\mu \nu}$.
Since Eq.(\ref{eq:Sigma_0}) 
is basically the same as the selfenergy $\Sigma_{\rm cen}$ in
the usual centrosymmetric systems, 
we consider $\Sigma_{\rm cen}$
for brevity.
The selfenergy at the hot spots for a sufficiently
clean system with the impurity damping $\gamma_{\rm imp}\ll v_F/\xi$
is calculated as
%%%%%%%%%%%%%%%%%%%%%%%%%%%%%%%%%%%%%
\begin{eqnarray}
{\rm Im}\Sigma_{\rm cen}^R(0,\vecc{k}_F)&=&g^2\sum_{q}
\int_{-\infty}^{\infty}\frac{d\varepsilon}{2\pi}
\Bigl[ \coth \frac{\varepsilon}{2T}-\tanh \frac{\varepsilon}{2T}\Bigr]
{\rm Im}\chi^R(\varepsilon,\vecc{q}){\rm Im}G^{0R}_{\rm cen}
(\varepsilon,
\vecc{k}_F-\vecc{q})\nonumber \\
&\sim&\sum_q\frac{\chi_Q\Gamma_s(\pi T)^2}{\omega_q(\omega_q+\pi T/2)}
{\rm Im}G^{0R}_{\rm cen}(0,\vecc{k}_F-\vecc{q})\nonumber \\
&\sim&\xi^0T\ln \bigl[ 1+\pi T/2\Gamma_s\bigr],
\end{eqnarray}
%%%%%%%%%%%%%%%%%%%%%%%%%%%%%%%%%%%%%%%
where $G^{0R}$ is the retarded Green's function including the impurity 
damping $\gamma_{\rm imp}$ and $\chi^R$ is the retarded susceptibility
obtained from the analytic continuation of Eq.(\ref{eq:chi}).
In the above calculation, the dispersion at the hot spots has been expanded as 
$\varepsilon_0(\vecc{k}_{F}-\vecc{q})=
\nabla \varepsilon_0(\vecc{k}_F-\vecc{Q})\cdot (\vecc{Q}-\vecc{q})$
for $\vecc{q}\simeq \vecc{Q}$,
because $\varepsilon_0(\vecc{k}_F-\vecc{Q})=-\varepsilon_0(\vecc{k}_F)=0$
is satisfied at the hot spots.
Here, we have used the approximation
$H(z)\equiv \int d\varepsilon [\coth (\varepsilon/2T)-
\tanh (\varepsilon/2T)]\varepsilon/(\omega_q^2+\varepsilon^2)
=1/z-2[\psi(z+1)-\psi(z+1/2)]\simeq (\pi T)^2/[\omega_q
(\omega_q+\pi T/2)]$, where $\psi$ is digamma function
and $z=\omega_q/(2\pi T)$.~\cite{pap:Kontani,pap:Stojkovic}
This approximate form becomes exact both for $z\rightarrow 0$ and
$z\rightarrow \infty$.
$\chi_Q,\Gamma_s$ and $\omega_q$ are defined as
$\chi_Q=\chi_0(\xi q_0)^2, \Gamma_s=\Gamma_0(\xi q_0)^{-2}$ and
$\omega_q=\Gamma_s[1+\xi^2(\vecc{q}-\vecc{Q})^2]$,
respectively.
Therefore, we have $\rho \sim -{\rm Im}\Sigma^{R} \sim T$
when the hot spots are dominant for the conductivity.
This is a general behavior for the clean 3D systems
with the 3D AF spin fluctuations.

In the numerical calculation, we neglect the real part of
the selfenergy which changes the shape of the Fermi surface,
because such an effect is non-perturbative.
We regard $\varepsilon_{\tau}(\vecc{k})$ as the dispersion 
that includes ${\rm Re}\Sigma_0$.
We show the numerical results for $\rho_{xx}=1/\sigma_{xx}$ 
by using Eq.(\ref{eq:Sigma_0})
and Eq.(\ref{eq:sigma_xx}) for clean limit.
%%%%%%%%%%%%%%%%%%%%%%%%%%%%%%%%%%%
\begin{figure}[htb]%
\begin{center}
\includegraphics[width=.3\linewidth,height=.225\linewidth]{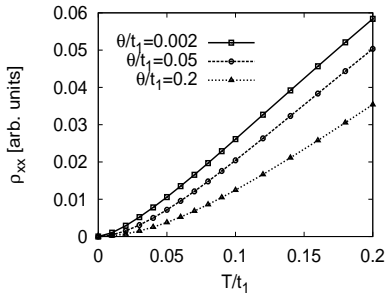}
\end{center}
\caption{%
Resistivity v.s. temperature for several $\theta$.
From the top to the bottom, $\theta/t_1=0.002,0.05,0.2$.
}
\label{fig:rho}
\end{figure}
%%%%%%%%%%%%%%%%%%%%%%%%%%%%%%%%%%%%
As shown in Fig.\ref{fig:rho}, for sufficiently small $\theta$,
the resistivity is proportional to $T$ in a wide range of
temperature where the hot spots are thermally blurred and
dominant for the conductivity.
In a very low temperature region where such blurring is suppressed,
$\rho$ is dominated by the electrons in the cold spots.
For large $\theta$, 
the canonical Fermi liquid behavior $\rho \sim T^2$
can be seen.
The calculated $\rho$ well explains the experimentally 
observed features of the resistivity in CeRhSi$_3$ and CeIrSi$_3$.
Therefore, we conclude that the observed $\rho \sim T$ above $T_c$
is due to the AF spin fluctuations.

We put a remark on the impurity effect.~\cite{pap:Rosch,pap:RoschPRB}
If the impurity scattering is sufficiently strong, 
the anisotropic scatterings by the spin fluctuations
are smeared, which weakens the singularity.
We, here, simply estimate the selfenergy by the spin fluctuations
$\Sigma_{\rm cen}^{\rm sf}$ in the presence of the impurities for 
centrosymmetric systems analytically.
For the system with the strong impurity effect which smears the anisotropy by
the AF spin fluctuations, 
we evaluate the selfenergy $\Sigma_{\rm cen}^{\rm sf}$
averaged on the Fermi surface
%%%%%%%%%%%%%%%%%%%%%%%%%%%%%%%%%%%%%
\begin{eqnarray}
\langle{\rm Im}\Sigma^{{\rm sf}R}_{\rm cen}
(0,\vecc{k})\rangle_{\rm FS}&\equiv&\frac{\sum_{k}
{\rm Im}\Sigma^{{\rm sf}R}_{\rm cen}(0,\vecc{k})
{\rm Im}G^{0R}_{\rm cen}(0,\vecc{k})}{
\sum_k{\rm Im}G^{0R}_{\rm cen}(0,\vecc{k})}
\nonumber \\
&\propto&\sum_q
\frac{\chi_Q\Gamma_s(\pi T)^2}{\omega_q(\omega_q+\pi T/2)}
\frac{\partial {\rm Im}\chi^{0R}_{\rm cen}(0,\vecc{q})}{\partial \omega}
\nonumber \\
&\sim&\xi^{-1}T\left(\sqrt{1+\pi T/2\Gamma_s}-1\right),
\end{eqnarray}
%%%%%%%%%%%%%%%%%%%%%%%%%%%%%%%%%%%%%%%
where we have defined $\chi^0_{\rm cen}(q)=-(T/N)\sum_k
G^0_{\rm cen}(k)G^0_{\rm cen}(k+q)$.
Here, we have assumed that its $T,q$-dependence is moderate and it does not 
contribute to the selfenergy.
We obtain $\rho \sim T^{3/2}$ for $\xi^{-2} \sim T$ in dirty
systems.
Note that, in the case of $\xi^{-2}\sim T^{3/2}$, 
we again have $\xi^{-1}T\left(\sqrt{1+\pi T/2\Gamma_s}-1\right)
\sim T^{3/4}T^1T^{-1/4}=T^{3/2}$ for sufficiently low temperatures.
Thus, the resistivity in the dirty systems with 3D AF spin fluctuations is
$\rho \sim T^{3/2}$ both for 
$\xi^{-2}\sim T$ and $\xi^{-2}\sim T^{3/2}$
in agreement with the previous studies.~\cite{pap:SCR1,pap:SCR2}

%%%%%%%%%%%%%%%%%%%%%%%%%%%%%%%%%%%%%%%%%%%%%%%%%%%%%
\section{Eliashberg equation in magnetic field}
\label{sec:Eliashberg}
%%%%%%%%%%%%%%%%%%%%%%%%%%%%%%%%%%%%%%%%%%%%%%%%%%%%%
%%%%%%%%%%%%%%%%%%%%%%%%%%%%%%%%
\subsection{exact formula within semiclassical approximation}
%%%%%%%%%%%%%%%%%%%%%%%%%%%%%%%
In this section, we derive a formula for the calculation of $H_{c2}$
from the linearized Eliashberg equation in real space.
The derivation is based on the semiclassical approximation which
is legitimate for the systems with $k_Fl_H\gg 1$, 
where $k_F$ is the Fermi wave
number and $l_H=1/\sqrt{|e|H}$ is the magnetic length.
This condition is satisfied for many superconductors including 
heavy fermion superconductors, and therefore, 
the resulting equation for $H_{c2}$
is applicable for a number of compounds.
Our formula is a generalization of the previous studies,~\cite{pap:WHH,
pap:Schossmann,pap:Bulaevskii}
and can be extended easily to more complicated models
although we use a single band model in this section.

To derive the formula for the calculation of $H_{c2}$, we use
the
linearized Eliashberg equation in real space
with the vector potential $\vecc{A}$ which gives a uniform magnetic field,
%%%%%%%%%%%%%%%%%%%%%%%%%%
\begin{eqnarray}
\Delta_{\alpha \alpha^{\prime}}(i\omega_n,\vecc{x},\vecc{x}^{\prime};\vecc{A})
&=&-\frac{1}{\beta}\sum_{i\omega_m}\sum_{\vecc{y}\vecc{y}^{\prime}}
V_{\alpha \alpha^{\prime},\beta \beta^{\prime}}
(i\omega_n,\vecc{x},\vecc{x}^{\prime};i\omega_m,\vecc{y},\vecc{y}^{\prime}
;\vecc{A})\nonumber \\
&&\times \sum_{\vecc{z}\vecc{z}^{\prime}}G_{\beta \gamma}
(i\omega_m,\vecc{y},\vecc{z};\vecc{A})
\Delta_{\gamma \gamma^{\prime}}(i\omega_m,\vecc{z},\vecc{z}^{\prime};\vecc{A})
G_{\beta^{\prime} \gamma^{\prime}}
(-i\omega_m,\vecc{y}^{\prime},\vecc{z}^{\prime};\vecc{A}),
\label{eq:Eliashberg_r}
\end{eqnarray}
%%%%%%%%%%%%%%%%%%%%%%%%
where $\sum_{x}$ represents the summation over all lattice sites,
and the spin indices are summed over.
$G,\Delta$ and $V$ are, respectively, the normal Green's function,
the gap function and the pairing interaction.
Note that, if $\vecc{A}$ is fully taken into account in the above
equation, the resulting equation is gauge invariant
under the gauge transformation $\vecc{A}(\vecc{x})
\rightarrow \vecc{A}(\vecc{x})+\nabla f(\vecc{x})$ and
$\psi(\tau,\vecc{x})\rightarrow \exp [ief(\vecc{x})]\psi(\tau,\vecc{x})$
where $\psi$ is the field operator of the electrons. 
By this transformation, each factor in the equation acquires the additional
phases as,
%%%%%%%%%%%%%%%%%%%%%%%%%%%
\begin{eqnarray}
G(i\omega_n,\vecc{x},\vecc{x}^{\prime})&\rightarrow &
\exp [ie\left(f(\vecc{x})-f(\vecc{x}^{\prime})
\right)]G(i\omega_n,\vecc{x},\vecc{x}^{\prime}),\\
\Delta(i\omega_n,\vecc{x},\vecc{x}^{\prime})&\rightarrow &
\exp [ie\left(f(\vecc{x})+f(\vecc{x}^{\prime})
\right)]\Delta(i\omega_n,\vecc{x},\vecc{x}^{\prime}),\\
V(i\omega_n,\vecc{x},\vecc{x}^{\prime};i\omega_m,\vecc{y},\vecc{y}^{\prime})
&\rightarrow &\exp [ie\left(
f(\vecc{x})+f(\vecc{x}^{\prime})-f(\vecc{y})-f(\vecc{y}^{\prime})\right)]
V(i\omega_n,\vecc{x},\vecc{x}^{\prime};i\omega_m,\vecc{y},\vecc{y}^{\prime}).
\end{eqnarray}
%%%%%%%%%%%%%%%%%%%%%%%%%%%%%
In this study, however, we use the semiclassical approximation in which
we do not explicitly include the effect of the
vector potential on the pairing interaction $V$, because 
the vector potential in $V$ is not responsible for the 
Landau quantization of the gap function which is the most important
phenomenon of the orbital effect in type-II superconductors.
By contrast, the lack of translational invariance
in $G$ and $\Delta$ in the presence of the applied vector potential 
$\vecc{A}$ is related to the Landau quantization.
Within the semiclassical approximation, the normal Green's function is 
%%%%%%%%%%%%%%%%%%%%%%%%%%%
\begin{eqnarray}
G(i\omega_n,\vecc{x},\vecc{y};\vecc{A})
&=&e^{i\varphi(\vecc{x},\vecc{y})}G(i\omega_n,\vecc{x}-\vecc{y}
;\vecc{A}=0), \\
\varphi(\vecc{x},\vecc{y})&=&e\int_{\vecc{y}}^{\vecc{x}}
\vecc{A}(\vecc{s})d\vecc{s}.
\label{eq:semicl}
\end{eqnarray}
%%%%%%%%%%%%%%%%%%%%%%%
We can easily perform the integral along the straight line 
$\vecc{s}(t)=\vecc{y}+t(\vecc{x}-\vecc{y}),0\leq t \leq 1$, using the relation
$\vecc{A}(a\vecc{x}+b\vecc{y})=a\vecc{A}(\vecc{x})+b\vecc{A}(\vecc{x})$
which holds for any $\vecc{A}$ giving a uniform magnetic field $\vecc{H}$,
and obtain
%%%%%%%%%%%%%%%%%%%%
\begin{eqnarray}
\varphi(\vecc{x},\vecc{y})=e\vecc{A}\left(\frac{\vecc{x}+\vecc{y}}{2}\right)
\cdot (\vecc{x}-\vecc{y}).
\end{eqnarray}
%%%%%%%%%%%%%%%%%%%%%%%
Although the linearized Eliashberg equation is
no longer invariant under the gauge transformation
defined above within this approximation,
it is still gauge invariant under a gauge transformation which
involves only the center of mass coordinate of the Cooper pairs.

Next, we proceed to rewrite Eq.(\ref{eq:Eliashberg_r}) in $k$-space.
The pairing interaction $V$ should be decomposed into two parts as,
%%%%%%%%%%%%%%%%%%%%%%
\begin{eqnarray}
V(i\omega_n,\vecc{x},\vecc{x}^{\prime};i\omega_m,\vecc{y},\vecc{y}^{\prime})
=V^{{\rm rel}}(i\omega_n,\vecc{x}-\vecc{x}^{\prime};i\omega_m,\vecc{y}-\vecc{y}^{\prime})\times
V^{{\rm cen}}\left(\frac{\vecc{x}+\vecc{x}^{\prime}}{2};
\frac{\vecc{y}+\vecc{y}^{\prime}}{2}\right)
\end{eqnarray}
%%%%%%%%%%%%%%%%%%%%%%
where $V^{{\rm rel}}$ and $V^{{\rm cen}}$ are the interactions
in the relative coordinate and the center of mass coordinate.
Here, we take $V^{{\rm cen}}$ to be dimensionless.
It is convenient to introduce the following variables
%%%%%%%%%%%%%%%%%%%%%%
\begin{align}
\begin{split}
&\vecc{R}=\frac{\vecc{x}+\vecc{x}^{\prime}}{2},\quad
\vecc{r}=\vecc{x}-\vecc{x}^{\prime},\\
&\vecc{Y}=\vecc{y}-\vecc{z},\quad \vecc{Y}^{\prime}=\vecc{y}^{\prime}-\vecc{z}^{\prime},\\
&\vecc{R}^{\prime}=\frac{\vecc{z}+\vecc{z}^{\prime}}{2},\quad
\vecc{r}^{\prime}=\vecc{z}-\vecc{z}^{\prime}.
\end{split}
\end{align}
%%%%%%%%%%%%%%%%%%%%%
In this coordinate, the phase factor $\exp (i\varphi(\vecc{y},\vecc{z})
+i\varphi(\vecc{y}^{\prime},\vecc{z}^{\prime}))$ which arises from 
$G(i\omega_m,\vecc{y},\vecc{z})
G(-i\omega_m,\vecc{y}^{\prime},\vecc{z}^{\prime})$ in 
Eq.(\ref{eq:Eliashberg_r}) becomes
%%%%%%%%%%%%%%%%%%%%%
\begin{eqnarray*}
\varphi(\vecc{y},\vecc{z})
+\varphi(\vecc{y}^{\prime},\vecc{z}^{\prime})
&=&e\vecc{A}(\vecc{R}^{\prime})(\vecc{Y}+\vecc{Y}^{\prime})
+e\vecc{A}(\vecc{r}^{\prime})(\vecc{Y}-\vecc{Y}^{\prime})
+e\vecc{A}(\vecc{Y})\vecc{Y}
+e\vecc{A}(\vecc{Y}^{\prime})\vecc{Y}^{\prime} \\
&\simeq&2e\vecc{A}(\vecc{R}^{\prime})\frac{\vecc{Y}+\vecc{Y}^{\prime}}{2}
+e\vecc{A}(\vecc{Y})\vecc{Y}
+e\vecc{A}(\vecc{Y}^{\prime})\vecc{Y}^{\prime}.
\end{eqnarray*}
%%%%%%%%%%%%%%%%%%%%%
In the second equality, the neglected term 
$e\vecc{A}(\vecc{r}^{\prime})(\vecc{Y}-\vecc{Y}^{\prime})$
is much smaller than the first
term, since for the dominant scattering processes,
$\|\vecc{Y}-\vecc{Y}^{\prime}\|,\|\vecc{r}^{\prime}\| \ll
\| \vecc{Y}+\vecc{Y}^{\prime}\|$ are satisfied
in the systems with short range pairing interaction.
The first term represents the phase which the Cooper pair with 
center of mass $\vecc{R}^{\prime}$ acquires.
We perform the Fourier transformation of Eq.(\ref{eq:Eliashberg_r}) and assume 
$V^{{\rm cen}}_{\alpha \alpha^{\prime},\beta \beta^{\prime}}
\left(\vecc{R};\vecc{R}^{\prime}+\frac{\vecc{Y}+\vecc{Y}^{\prime}}{2}\right)
=\delta_{R,R^{\prime}+(Y+Y^{\prime})/2}$,
then we obtain
%%%%%%%%%%%%%%%%%%%%%
\begin{eqnarray*}
\Delta_{\alpha \alpha^{\prime}}(i\omega_n,\vecc{r},\vecc{R})
&=&-\frac{1}{\beta}\sum_{i\omega_m}
\frac{1}{N^2}\sum_{\vecc{k}\vecc{k}^{\prime}}
\frac{1}{N^2}\sum_{\vecc{p}\vecc{p}^{\prime}}
\sum_{\vecc{Y}\vecc{Y}^{\prime}}
V^{{\rm rel}}_{\alpha \alpha^{\prime},\beta \beta^{\prime}}
(i\omega_n,\vecc{k};i\omega_m,\vecc{k}^{\prime})\\
&&\times G_{\beta \gamma}(i\omega_m,\vecc{p})
G_{\beta^{\prime} \gamma^{\prime}}
(-i\omega_m,\vecc{p}^{\prime})
\Delta_{\gamma \gamma^{\prime}}
\left(i\omega_m,\vecc{k}^{\prime},
\vecc{R}-\frac{\vecc{Y}+\vecc{Y}^{\prime}}{2}\right)\\
&&\times \exp i\left(\vecc{k}\vecc{r}+(-\vecc{k}^{\prime}+\vecc{p})\vecc{Y}
+(\vecc{k}^{\prime}+\vecc{p}^{\prime})\vecc{Y}^{\prime}\right)\\
&&\times \exp i\left(e\vecc{A}(\vecc{Y})\vecc{Y}
+e\vecc{A}(\vecc{Y}^{\prime})\vecc{Y}^{\prime}
+e\vecc{A}\left(\vecc{R}-\frac{\vecc{Y}+\vecc{Y}^{\prime}}{2}\right)
(\vecc{Y}+\vecc{Y}^{\prime}) \right).
\end{eqnarray*}
%%%%%%%%%%%%%%%%%%%%%%%%
The phase factor including $\vecc{A}$ is rewritten as
%%%%%%%%%%%%%%%%%%%%%%%%
\begin{eqnarray*}
&&\exp i\left(e\vecc{A}(\vecc{Y})\vecc{Y}
+e\vecc{A}(\vecc{Y}^{\prime})\vecc{Y}^{\prime}
+2e\vecc{A}\left(\vecc{R}-\frac{\vecc{Y}+\vecc{Y}^{\prime}}{2}\right)
\frac{\vecc{Y}+\vecc{Y}^{\prime}}{2} \right)
\Delta_{\gamma \gamma^{\prime}}
\left(i\omega_m,\vecc{k}^{\prime},
\vecc{R}-\frac{\vecc{Y}+\vecc{Y}^{\prime}}{2}\right)\\
&&=e^{i\theta_1+i\theta_2}
\exp i\left(
-\frac{\vecc{Y}+\vecc{Y}^{\prime}}{2}\vecc{\Pi}(\vecc{R}) \right)
\Delta_{\gamma \gamma^{\prime}}
\left(i\omega_m,\vecc{k}^{\prime},
\vecc{R}\right)
\end{eqnarray*}
%%%%%%%%%%%%%%%%%%%%%%%
where $\vecc{\Pi}(\vecc{R})=-i\nabla -2e\vecc{A}(\vecc{R})$,
$\theta_1=-e[(\vecc{Y}+\vecc{Y}^{\prime})/2]^2\nabla_{R}
\vecc{A}(\vecc{R})$ and $\theta_2=(e/2)
\vecc{A}(\vecc{Y}-\vecc{Y}^{\prime})(\vecc{Y}-\vecc{Y}^{\prime})$.
$\theta_2$ is proportional to $(\vecc{Y}-\vecc{Y}^{\prime})$, and
therefore, negligible.
$\theta_1$ is also small compared with $\vecc{\Pi}
(\vecc{Y}+\vecc{Y}^{\prime})/2$, because $\theta_1\sim 1/(k_Fl_H)^2$
while $\vecc{\Pi}(\vecc{Y}+\vecc{Y}^{\prime})/2\sim 1/(k_Fl_H)$.
Neglecting $\theta_1$ and $\theta_2$,
we end up with the linearized Eliashberg equation
in $k$-space in the presence of the vector potential $\vecc{A}$,
%%%%%%%%%%%%%%%%%%%%%%%%%%
\begin{eqnarray}
\Delta_{\alpha \alpha^{\prime}}(k,\vecc{R})
&=&-\frac{1}{\beta N}\sum_{k^{\prime}}
V_{\alpha \alpha^{\prime},\beta \beta^{\prime}}
(k,k^{\prime})G_{\beta \gamma}(k^{\prime}+\Pi/2)
G_{\beta^{\prime} \gamma^{\prime}}(-k^{\prime}+\Pi/2)
\Delta_{\gamma \gamma^{\prime}}(k^{\prime},\vecc{R}),
\label{eq:Eliashberg_k}
\end{eqnarray}
%%%%%%%%%%%%%%%%%%%%%%%%%%%
where $k=(i\omega_n,\vecc{k})$ and $\Pi=(0,\vecc{\Pi})$,
and we have written $V^{\rm rel}$ as $V$ for simplicity.
This is a well-known form of the Eliahsberg equation and similar expressions
are often used for the discussion of $H_{c2}$ in superconductors.
As mentioned before, if we define a semiclassical gauge transformation
which involves only $\vecc{R}$ as 
%%%%%%%%%%%%%%%%%%%%%%%%%%%%
\begin{eqnarray}
\Delta(k,\vecc{R})\rightarrow
\exp[i2ef(\vecc{R})]\Delta(k,\vecc{R}),
\end{eqnarray}
%%%%%%%%%%%%%%%%%%%%%%%%%%
this equation is gauge invariant,
because, for ${\mathcal O}[\vecc{A}(\vecc{R})]\equiv
G(k+\Pi/2)G(-k+\Pi/2)$, 
$e^{-i2ef}{\mathcal O}[\vecc{A}+\nabla
f]e^{i2ef}={\mathcal O}[\vecc{A}]$ is satisfied.
The relative coordinate is not involved in the gauge transformation
in the semiclassical approximation, and $V^{\rm rel}(k,k^{\prime})$ does not 
change under the transformation.

We, next, proceed to rewrite the above Eliashberg equation to
perform numerical calculations.
In the present study, we denote the coordinate as 
$(R_1,R_2,R_3)=(R_x,R_y,R_z)$ 
for the perpendicular field and $(R_1,R_2,R_3)=(R_x,R_z,R_y)$
for the in-plane field.
With this notation,
the gap function for $\vecc{H}=H\vecc{e}_3$ 
is expanded by the Landau functions,
%%%%%%%%%%%%%%%%%%%%%%%%%%%%%%
\begin{eqnarray}
\Delta_{\alpha \alpha^{\prime}}(k,\vecc{R})&=&\sum_{n=0}^{\infty}
 \Delta_{\alpha \alpha^{\prime}n}(k)\phi_{\Lambda Qn}(\vecc{R}),\\
\phi_{\Lambda Qn}(\vecc{R})&=&e^{i\vecc{Q}^{\Lambda}\vecc{R}^{\Lambda}}
\phi_n(R_1^{\Lambda},
R_2^{\Lambda}),
\end{eqnarray}
%%%%%%%%%%%%%%%%%%%%%%%%%%%%
where $\{\phi_n\}$ are the usual Landau functions,
$\vecc{R}^{\Lambda}=(\Lambda^{1/2}R_1,\Lambda^{-1/2}R_2,R_3)$,
and $\vecc{Q}^{\Lambda}=(\Lambda^{-1/2}Q_1,\Lambda^{1/2}Q_2,Q_3)$.
The parameters $\vecc{Q}$ and $\Lambda$ represent,
respectively, the modulation of the gap function
and the anisotropy of the vortex lattice in the $R_1R_2$ plane, and
both of them are optimized to give the largest $H_{c2}$.
We introduce the operator
$\vecc{\Pi}^{\Lambda Q}=\left( -i\nabla^{\Lambda}-2e\vecc{A}
(\vecc{R}^{\Lambda})\right)-\vecc{Q}^{\Lambda}$
with
$\nabla^{\Lambda}=\partial /\partial \vecc{R}^{\Lambda}$.
The Landau functions $\{ \phi_{\Lambda Qn}\}$ 
satisfy the following relations,
%%%%%%%%%%%%%%%%%%%%%%%%%%%%%%%%%%
\begin{eqnarray}
\Pi_{+}^{\Lambda Q}\phi_{\Lambda Qn}&=&\sqrt{n+1}\phi_{\Lambda Qn+1},\\
\Pi_{-}^{\Lambda Q}\phi_{\Lambda Qn}&=&\sqrt{n}\phi_{\Lambda Qn-1}
\end{eqnarray}
%%%%%%%%%%%%%%%%%%%%%%%%%%%%%%%%%%
where $\Pi_{\pm}^{\Lambda Q}=
\frac{l_H}{2}\left(\Pi_{1}^{\Lambda Q}\mp i\Pi_2^{\Lambda Q}\right)$.

By taking an inner product of Eq.(\ref{eq:Eliashberg_k}), we obtain
%%%%%%%%%%%%%%%%%%%%%%%%%%%%%%
\begin{eqnarray}
\Delta_{\alpha \alpha^{\prime}n}(k)&=&-\frac{T}{N}\sum_{k^{\prime}}
 V_{\alpha \alpha^{\prime}\beta \beta^{\prime}}(k,k^{\prime})
\sum_{\tau \tau^{\prime}}l_{\tau \beta \gamma}(k^{\prime})
 l_{\tau^{\prime} \beta^{\prime} \gamma^{\prime}}(-k^{\prime})
\tilde{{\mathcal G}}_{\tau \tau^{\prime}nn^{\prime}}(k^{\prime})
\Delta_{\gamma \gamma^{\prime}}(k^{\prime},\vecc{R}),
\label{eq:Delta_pi/2}\\
\tilde{{\mathcal G}}_{\tau_1 \tau_2n_1n_2}(k,H)&=&
\sum_{m=0}^{\infty}\langle \phi_{\Lambda Q n_1}
|G_{\tau_1}(k+\Pi/2)|\phi_{\Lambda Q m}\rangle \langle
\phi_{\Lambda Q m}| G_{\tau_2}(-k+\Pi/2)|\phi_{\Lambda Q n_2}\rangle ,
\label{eq:G_pi/2}
\end{eqnarray}
%%%%%%%%%%%%%%%%%%%%%%%%%%%%%%%%%%%%%%%%%
where the completeness relation $\sum_m|\phi_{\Lambda Q m}\rangle \langle
\phi_{\Lambda Q m}|=1$ is used.
Here, we have neglected the $\Pi$ operators in $l_{\tau}$ because
they only lead to the terms with positive powers of $eH$, i.e.,
$l_{\tau}(k+\Pi)\simeq l_{\tau}(k)
+\nabla l_{\tau}(k)\vecc{\Pi}
={\mathcal O}(1)+{\mathcal O}(1/k_Fl_H)$, while $\tilde{{\mathcal G}}$ is 
proportional to $(|e|H)^{-1/2}$ describing the non-perturbative effect of
the formation of the vortex lattice.

Within the semiclassical approximation, 
Eqs. (\ref{eq:Delta_pi/2}) and (\ref{eq:G_pi/2}) are exact.
For numerical calculations, however, we need a cut off 
in the summation $\sum_{m=0}^{\infty}$
which should be large
enough for the calculated results to be reliable.
It is hard to
solve Eq.(\ref{eq:Delta_pi/2}) and (\ref{eq:G_pi/2}) with such a large cut off.
So, we introduce an alternative formula for the numerical calculation
of $H_{c2}$ in the next section.

%%%%%%%%%%%%%%%%%%%%%%%%%%%%%%%%%%%%%%%%
\subsection{alternative formula for numerical calculation}
%%%%%%%%%%%%%%%%%%%%%%%%%%%%%%%%%%%%%%%%
As mentioned at the end of the previous section,
it is difficult to solve
the exact formula Eqs.(\ref{eq:Delta_pi/2}) and (\ref{eq:G_pi/2})
numerically.
Then, we approximate them by an alternative equation.
Instead of Eq.(\ref{eq:Eliashberg_k}),
we introduce a modified Eliashberg equation,
%%%%%%%%%%%%%%%%%%%%%%%%%%%%%%%%
\begin{eqnarray}
\Delta_{\alpha \alpha^{\prime}}(k,\vecc{R})
=-\frac{T}{2N}\sum_{k^{\prime}}
 V_{\alpha \alpha^{\prime}\beta \beta^{\prime}}(k,k^{\prime})
[ G_{\beta \gamma}(k^{\prime}+\Pi)G_{\beta^{\prime}\gamma^{\prime}}
(-k^{\prime})
 +G_{\beta \gamma}(k^{\prime})G_{\beta^{\prime}\gamma^{\prime}}
(-k^{\prime}+\Pi)]
\Delta_{\gamma \gamma^{\prime}}(k^{\prime},\vecc{R}).
\label{eq:Eliashberg_kpi}
\end{eqnarray}
%%%%%%%%%%%%%%%%%%%%%%%%%%%%%%%%
This equation is rewritten as
%%%%%%%%%%%%%%%%%%%%%%%%%%%%%%%
\begin{eqnarray}
\Delta_{\alpha \alpha^{\prime}n}(k,\vecc{R})&=&-\frac{T}{N}\sum_{k^{\prime}}
 V_{\alpha \alpha^{\prime}\beta \beta^{\prime}}(k,k^{\prime})
\sum_{\tau \tau^{\prime}}l_{\tau \beta \gamma}(k^{\prime},H)
 l_{\tau^{\prime} \beta^{\prime} \gamma^{\prime}}(-k^{\prime},H)
{\mathcal G}_{\tau \tau^{\prime}nn^{\prime}}(k^{\prime},H)
\Delta_{\gamma \gamma^{\prime}}(k^{\prime},\vecc{R}),
\label{eq:Delta_pi}\\
{\mathcal G}_{\tau_1 \tau_2n_1n_2}(k,H)&=&
\frac{1}{2}[\langle \phi_{\Lambda Q n_1}
|G_{\tau_1}(k+\Pi)|\phi_{\Lambda Q n_2}\rangle G_{\tau_2}(-k)+
G_{\tau_1}(k)\langle
\phi_{\Lambda Q n_1}| G_{\tau_2}(-k+\Pi)|\phi_{\Lambda Q n_2}\rangle ].
\label{eq:G_pi}
\end{eqnarray}
%%%%%%%%%%%%%%%%%%%%%%%%%%%%%%%%
This is the alternative formula for numerical calculations,
and does not need an infinite summation like $\sum_{m}$ in
Eq.(\ref{eq:G_pi/2}).
Equation(\ref{eq:Eliashberg_k}) is not exactly equivalent to
Eq.(\ref{eq:Eliashberg_kpi}) when $V$ and $\Delta$ are $k$-dependent
as in unconventional superconductors. However,
we have confirmed that the two different formulae give 
the qualitatively same results for $H_{c2}$, and the
quantitative difference is small.
Therefore, hereafter, we use Eqs.(\ref{eq:Delta_pi}) and
(\ref{eq:G_pi}).

With the use of the relation 
$\frac{1}{a}=\int_0^{\infty}dte^{-at}$ for Re$(a)>0$, the matrix elements are
calculated as
%%%%%%%%%%%%%%%%%%%%
\begin{eqnarray}
\langle \phi_{\Lambda Q n_1}
|G_{\tau_1}(k+\Pi)|\phi_{\Lambda Q n_2}\rangle
&=&-is_{\tau_1}\sum_{l=0}^{{\rm min}\{n_1,n_2\}}\frac{\sqrt{n_1!n_2!}}
{(n_1-l)!(n_2-l)!l_1!}c_{1+}^{n_1-l}c_{1-}^{n_2-l}
\int_0^{\infty}dte^{-\frac{a_1}{2}t^2-b_1t}t^{n_1+n_2-2l}\nonumber \\
&=&-is_{\tau_1}\sum_{l=0}^{{\rm min}\{n_1,n_2\}}\frac{\sqrt{n_1!n_2!}}
{(n_1-l)!(n_2-l)!l_1!}c_{1+}^{n_1-l}c_{1-}^{n_2-l}
\left( \frac{2}{a_1}\right)^{\frac{n_1+n_2-2l+1}{2}}
F_{n_1+n_2-2l}(z_1),\label{eq:G_pi+} \\
\langle \phi_{\Lambda Q n_1}
|G_{\tau_2}(-k+\Pi)|\phi_{\Lambda Q n_2}\rangle
&=&is_{\tau_2}\sum_{l=0}^{{\rm min}\{n_1,n_2\}}\frac{\sqrt{n_1!n_2!}}
{(n_1-l)!(n_2-l)!l_1!}c_{2+}^{n_1-l}c_{2-}^{n_2-l}
\left( \frac{2}{a_2}\right)^{\frac{n_1+n_2-2l+1}{2}}
F_{n_1+n_2-2l}(z_2)\label{eq:G_pi-}
\end{eqnarray}
%%%%%%%%%%%%%%%%%%%%%%%%%%
where 
%%%%%%%%%%%%%%%%%%%%%%%%%
\begin{eqnarray}
F_N(z)&=&\sum_{n=0}^N\binom{N}{n}(-z)^{N-n}f_n(z),\\
f_n(z)&=&e^{z^2}\int_z^{\infty}dte^{-t^2}t^n.
\label{eq:f_def}
\end{eqnarray}
%%%%%%%%%%%%%%%%%%%%%%%%%%%%%
The variables $a,b,c$ and $z$ are given by,
%%%%%%%%%%%%%%%%%%%%%%%%%%%%%%
\begin{align}
\begin{split}
a_1&=A_{\tau_1}(k;H)=l_H^{-2}[v_{\tau_1 \perp}^{\Lambda}
(k;H)]^2\\
b_1&=s_{\tau_1}\left(\omega_{\tau_1}(k;H)+i[\tilde{\varepsilon}_{\tau_1}
(k;H)+\vecc{v}_{\tau_1}^{\Lambda}(k;H)\cdot \vecc{Q}^{\Lambda}]\right),\\
z_1&=b_1/\sqrt{2a_1},\\
c_{1\pm}&=C_{\tau_1\pm}(k;H)=
-is_{\tau_1}l_H^{-1}v_{\tau_1\pm}^{\Lambda}(k;H),\\
s_{\tau_1}&={\rm sgn}(\omega_{\tau_1}(k;H)),\\
v_{\tau \pm}^{\Lambda}(k;H)&=v_{\tau 1}^{\Lambda}(k;H)\pm 
iv_{\tau 2}^{\Lambda}(k;H),\\
v_{\tau \perp}^{\Lambda}&=\sqrt{v_{\tau +}^{\Lambda}v_{\tau -}^{\Lambda}},
\end{split}
\end{align}
%%%%%%%%%%%%%%%%%%%%%%%%%%%%%
and 
%%%%%%%%%%%%%%%%%%%%%%%%%%%%%%
\begin{align}
\begin{split}
a_2&=A_{\tau_2}(k;-H),\\
b_2&=s_{\tau_2}\left(\omega_{\tau_2}(k;-H)+i[-\tilde{\varepsilon}_{\tau_2}
(k;-H)+\vecc{v}_{\tau_2}^{\Lambda}(k;-H)\cdot \vecc{Q}^{\Lambda}]\right),\\
z_2&=b_2/\sqrt{2a_2},\\
c_{2\pm}&=C_{\tau_2\pm}(k;-H),\\
s_{\tau_2}&={\rm sgn}(\omega_{\tau_2}(k;-H)),
\end{split}
\end{align}
%%%%%%%%%%%%%%%%%%%%%%%%%%%%%
where $\omega_{\tau}(k;H)=\omega_n-{\rm Im}\Sigma_{0}(k,H)$,
$\tilde{\varepsilon}_{\tau}(k,H)=\varepsilon_{\tau}(k,H)
+{\rm Re}\Sigma_0(k,H)$, and
$\vecc{v}^{\Lambda}(k,H)=\nabla^{\Lambda}\tilde{\varepsilon}_{\tau}(k,H)$.

A convenient expression of $f_n$ is obtained 
through the recurrence formula which is directly derived from
Eq.(\ref{eq:f_def}),
%%%%%%%%%%%%%%%%%%%%%%%%%%%%%%%%%
\begin{eqnarray}
&&f_0(z)=\frac{\sqrt{\pi}}{2}e^{z^2}{\rm erfc}(z),\\
&&f_1(z)=\frac{1}{2},\\
&&f_n(z)-\frac{n-1}{2}f_{n-2}(z)-\frac{1}{2}z^{n-1}=0\qquad (n\geq 2).
\end{eqnarray}
%%%%%%%%%%%%%%%%%%%%%%%%%%%%%%%%
The solution is
%%%%%%%%%%%%%%%%%%%%%%%%%%%%%%%%%%%%
\begin{eqnarray}
f_n(z)&=&\frac{(n-1)!!}{2^{q_n}}f_{r_n}(z)+\sum_{k=1}^{q_n}
\frac{(n-1)!!}{2^k(n-2k+1)!!}z^{n-2k+1},
\end{eqnarray}
%%%%%%%%%%%%%%%%%%%%%%%%%%%%%%
where $q_n=\frac{n}{2}(n:\mbox{even}),\frac{n-1}{2}(n:\mbox{odd})$
and $r_n=0(n:\mbox{even}),1(n:\mbox{odd})$.

With the expression Eqs.(\ref{eq:G_pi+}) and (\ref{eq:G_pi-}),
the numerical calculation of Eq.(\ref{eq:Delta_pi}) is straightforward.
Similar expression for $\tilde{\mathcal G}$ can be obtained in the 
same way and we can also solve Eq.(\ref{eq:Delta_pi/2}) numerically.
As mentioned above, 
Eq.(\ref{eq:Delta_pi/2}) 
and Eq.(\ref{eq:Delta_pi}) give the qualitatively same
results and the quantitative difference is small.

The important point is that
Eq.(\ref{eq:Delta_pi}) 
allows us to calculate $H_{c2}$ for general lattice models with
arbitrary Fermi surfaces,
taking into account both the orbital and the Pauli depairing effect on 
an equal footing. 
In the Gintzburg-Landau approach, the relative strength of the 
orbital and the Pauli depairing effect is characterized by the Maki parameter 
$\alpha_M=\sqrt{2}H_{\rm orb}/H_{\rm P}$, where $H_{\rm orb}$ and
$H_{\rm P}$ are the orbital and the Pauli limiting field,
respectively.
In our formulation, however, we do not need such a parameter which
is difficult to be determined experimentally.
The parameter corresponding to $\alpha_M$ in this study is 
an effective mass of the quasiparticle 
for the cyclotron motion
%%%%%%%%%%%%%%%%%%%%%%%%%%%%%%%
\begin{eqnarray}
m_{\rm eff}=\frac{\hbar^2}{t_1a^2},
\end{eqnarray}
%%%%%%%%%%%%%%%%%%%%%%%%%%%%%%
where $t_1$ is the energy unit of the lattice model and $a$ is 
the length unit, i.e., the lattice constant.
Writing $l_H=\tilde{l}_Ha$ and $\mu_BH=
\tilde{h}t_1$ with dimensionless variables $\tilde{l}_H$ and $\tilde{h}$,
we have a simple identity,
%%%%%%%%%%%%%%%%%%%%%%%%%%%%%%%
\begin{eqnarray}
\tilde{l}^{-2}_{H}&=&\frac{|e|\hbar}{\mu_B}\frac{\tilde{h}}{m_{\rm eff}}.
\end{eqnarray}
%%%%%%%%%%%%%%%%%%%%%%%%%%%%%%
A large effective mass corresponds to a slow velocity of 
the quasiparticles for the cyclotron motion leading to a suppression
of the orbital depairing effect.
%The difference between $m_{\rm eff}$ and $\alpha_M$ is that
$m_{\rm eff}$ can be determined reasonably, while evaluating $\alpha_M$
from experiments is rather difficult because 
$H_{\rm orb}$ and $H_{\rm P}$ are not directly observed, 
especially for the strong coupling superconductors.
The lattice constant is determined experimentally, and 
we fix $a=4.0$(\AA) in this study, which is consistent with the 
experiments.~\cite{pap:Muro,pap:Okuda}
We can also determine the value of $t_1$ in a reasonable way.
By solving the Eliashberg equation(\ref{eq:Eliashberg_kpi}) at $H=0$,
we obtain $T_c=\tilde{T}_ct_1$ in the unit of $t_1$.
Then, comparing it with the experimentally observed transition temperature
$T_c^{\rm exp}$, we have
%%%%%%%%%%%%%%%%%%%%%%%%%%%%%%%
\begin{eqnarray}
t_1&=&\frac{T_c^{\rm exp}}{\tilde{T}_c}\quad ({\rm K}).
\label{eq:t_1}
\end{eqnarray}
%%%%%%%%%%%%%%%%%%%%%%%%%%%%%%
In this way, the parameters of the model are evaluated.
However, the choice of all the parameters is not unique and 
there remains some ambiguity especially for the strength of the
interaction on which $T_c=\tilde{T}_ct_1$ largely depends.
Therefore, we change the value of the strength of the 
interaction depending on the choice of the magnitude of $t_1$
to make $T_c$ consistent with the observed values.
For the calculation of $H_{c2}$ in CeRhSi$_3$ and
CeIrSi$_3$, we use two values of $t_1$ and compare the results.

In addition to the treatment of the Pauli and the orbital depairing effect, 
the strong coupling effect can be
included naturally in Eq.(\ref{eq:Delta_pi}).
Once we calculate the pairing interaction $V$ and the selfenergy
$\Sigma$ for a given Hamiltonian, they are directly incorporated into the
Eliashberg equation(\ref{eq:Delta_pi}).
This feature 
is essentially important for the study of 
$H_{c2}$ in CeRhSi$_3$ and CeIrSi$_3$, 
because it is considered that they are located near the
AF QCPs and the quasiparticles interact with each other
through the strong spin fluctuations.

%%%%%%%%%%%%%%%%%%%%%%%%%%%%%%%%%%%%%%%%%%%%%%%%%%%
\section{Calculation of Upper Critical Field}
\label{sec:Hc2}
%%%%%%%%%%%%%%%%%%%%%%%%%%%%%%%%%%%%%%%%%%%%%%%%%%%
In this section, we show the numerical results
calculated from Eq.(\ref{eq:Delta_pi}) with
Eq.(\ref{eq:G_pi}).
We solve the Eliashberg equation both for $H=0$ and $H\neq 0$.
For the latter, we study the two cases:
$H=(0,0,H)$ and $H=(0,H,0)$.
In the case of $H\parallel \hat{z}$, the Pauli depairing effect is 
strongly suppressed by the anisotropic spin-orbit interaction,
and $H_{c2}$ is determined by the orbital limiting field
$H_{\rm orb}$.~\cite{pap:Frigeri,pap:FujimotoJPSJ1,pap:FujimotoJPSJ2}
On the other hand, for $H\perp \hat{z}$, the Pauli depairing effect is
significant because of the anisotropic distortion of the Fermi surface
due to the Rashba SO interaction,
and $H_{c2}$ is mainly determined by the Pauli limiting field $H_{\rm P}$.

In this section, we use the action Eq.(\ref{eq:action}) 
to calculate $H_{c2}$.
The selfenergy $\Sigma_0$ has the real and the imaginary part which have
different effects, respectively.
${\rm Re}\Sigma_0$ only gives the deformation of the Fermi surface,
and, as in Sec.\ref{sec:normal},
it is reasonable to consider that $\varepsilon_{\tau}(\vecc{k})$ already
includes the shift due to ${\rm Re}\Sigma_0$ and to replace 
$\varepsilon_{\tau}(\vecc{k})+{\rm Re}\Sigma_0
\rightarrow \varepsilon_{\tau}(\vecc{k})$.
On the other hand, ${\rm Im}\Sigma_0(i\omega_n,\vecc{k})$ 
gives two important effects for the
quasiparticles around the Fermi level.
One is the damping factor $\gamma=-{\rm Im}\Sigma_0^R$ and
the other is the mass enhancement factor $z^{-1}=[1-
\partial {\rm Re}\Sigma_0^R(0)/\partial \omega]$.
Especially, the former gives rise to the depairing effects of 
the Cooper pair due to the inelastic scattering,
which would lower $T_c$. 
For $T\ll T_c$, however, such a suppression 
does not occur because ${\rm Im}\Sigma_0\rightarrow 0$ as $T\rightarrow 0$.
This property is a key for the colossal enhancement in $H_{c2}$ for
$H\parallel c$-axis.

We, next, consider the pairing interaction between the quasiparticles
due to the strong spin fluctuations near the QCP.
They are evaluated at the lowest order in $g^2\chi_0$,
%%%%%%%%%%%%%%%%%%%%%%%%%%%%%%%%%
\begin{eqnarray}
V_{ss,ss}(k,k^{\prime})&=&-\frac{1}{6}g^2
       \chi(k-k^{\prime})+\frac{1}{6}g^2\chi(k+k^{\prime}),\label{eq:int_1}\\
V_{s\bar{s},s\bar{s}}(k,k^{\prime})&=&
      \frac{1}{6}g^2\chi(k-k^{\prime})+\frac{1}{3}g^2\chi(k+k^{\prime}),
\label{eq:int_2}\\
V_{s\bar{s},\bar{s}s}(k,k^{\prime})&=&-V_{s\bar{s},s\bar{s}}(k,-k^{\prime}),
\label{eq:int_3}
\end{eqnarray}
%%%%%%%%%%%%%%%%%%%%%%%%%%%%%%%%%%%%
and the other components 
are zero. These are directly derived from Eq.(\ref{eq:action_SF}).
Although the applied fields might affect $V$, 
we neglect such an effect in this section.
The $H$-dependence of $V$ can be included within our approach,
if the field dependence of $\chi(q)$ is clarified by some experiments.
We also note that, in Eqs.(\ref{eq:int_1})$\sim$(\ref{eq:int_3}), 
the spin-flip 
scattering processes are not included.
They are expected to enhance the mixing of the spin singlet and the triplet
superconductivity.
These two neglected effects are discussed in Sec.\ref{sec:2points}.
As noted in Sec.\ref{sec:normal}, 
the coupling constant $g$ should be regarded
as an effective one renormalized
by the vertex corrections~\cite{pap:Yonemitsu,pap:Monthoux_ver}.

%%%%%%%%%%%%%%%%%%%%%%%%%%%%%%%%%%%%%%%%%%
\subsection{$H$=0 case}
\label{subsec:H=0}
%%%%%%%%%%%%%%%%%%%%%%%%%%%%%%%%%%%%%%%%%%%%
In this section,
we study the gap function and the
transition temperature at $H=0$ by solving Eq.(\ref{eq:Eliashberg_kpi}).
In this case, $\Delta$ does not depend on the center of mass $\vecc{R}$
and ${\mathcal G}$ is simplified as
${\mathcal G}_{\tau \tau^{\prime}}(k)=G_{\tau}(k)G_{\tau^{\prime}}(-k)$.
Among the five irreducible representations of the point group C$_{4v}$ for
CeRhSi$_3$ and CeIrSi$_3$, 
the most stable symmetry of the gap functions is A$_1$ symmetry which
is consistent with the previous study.~\cite{pap:TadaJPSJ}
The $k$-dependence of the singlet gap function is
$\Delta_{\rm singlet}\sim \cos(2k_za)$, and that of the 
triplet gap function is $\Delta_{\rm triplet}\sim \sin(k_{x,y})$,
as will be discussed in detail in Sec.\ref{subsec:flipping}.
In the previous study for $H_{c2}\parallel \hat{z}$,
we have neglected the triplet part of the gap function,
because it is much smaller than the singlet one 
in amplitude.~\cite{pap:TadaPRL}
In the present study, we take it into account and show
that the results in the previous study are not changed.

\begin{figure}[htb]%
\begin{center}
\includegraphics[width=.3\linewidth,height=.225\linewidth]{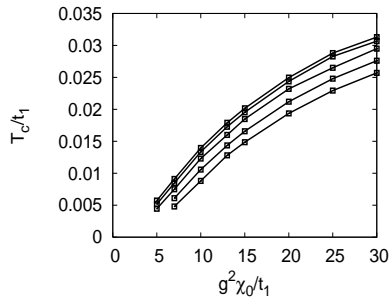}
\end{center}
\caption{%
Transition temperatures $T_c/t_1$ as functions of $g^2\chi_0/t_1$
for several $\theta$ at $H=0$.
The curves correspond to $\theta/t_1=0.002,0.005,0.01,0.02$ and $0.03$
from the top to the bottom.
}
\label{fig:g-t}
\end{figure}

In Fig.(\ref{fig:g-t}), the transition temperatures for this A$_1$ symmetric
superconducting state for several $\theta$
are shown as functions of $g^2\chi_0$. 
%%%%%%%%%%%%%%%%%%%%%%%%%%%%%%%%%%%
%%%%%%%%%%%%%%%%%%%%%%%%%%%%%%%%%%%%
$T_c$ saturates for large $g^2\chi_0$ because 
the strength of the pairing interaction and that of the 
depairing effect through the normal selfenergy become comparable.
Note that the dependence of $T_c$ on $\theta$ is weak.

The coupling constant $g$ is fixed so that the calculated $T_c$ is
of the same order as the experimentally observed $T_c$.
In CeRhSi$_3$ and CeIrSi$_3$, Kondo temperature is 
$T_{\rm K}\sim$ 50-100 (K)~\cite{pap:Muro98}
and the resistivity saturates around 
200$\sim$300 (K),~\cite{pap:Okuda,pap:KimuraJPSJ}
which implies that
the hopping integral $t_1$ in our model is $t_1\sim 50$-100 (K).
On the other hand, observed $T_c$ is $T_c\sim 1$(K), that is,
$T_c\sim 0.01t_1$-$0.02t_1$.
To reproduce this $T_c$ in the calculation, we fix $g^2\chi_0/t_1=10$-15.
For these values, the system is in a strong coupling region.
The renormalization factor averaged
on the Fermi surface is $z^{-1}\sim 1.7$
for $g^2\chi_0/t_1=10$, and 
$z$ is not sensitive to $\theta$, which is characteristic of 
the 3D AF spin fluctuations.~\cite{pap:SCR1,pap:SCR2,pap:Hertz,pap:Millis} 
Below, we mainly study the case of $g^2\chi_0/t_1=10$ for which
$T_c$ for the minimum $\theta =0.002t_1$ is $T_c=0.0139t_1$.
Setting $T_{c}=1.3$(K), 
which is an averaged value of $T_c$ for CeRhSi$_3$~\cite{pap:Kimura} 
and CeIrSi$_3$,~\cite{pap:Sugitani}
we have $t_1=93.8$(K).
We also consider the case of $g^2\chi_0/t_1=15$ for in-plane fields
in Sec.\ref{subsec:Hab}.
In this case, similarly, we have $T_c=0.0199t_1\equiv 1.3$ (K) 
and $t_1=65.3$(K).

%%%%%%%%%%%%%%%%%%%%%%%%%%%%%%%%%%%%%%%%%%%%
\subsection{$H\parallel c$-axis case}
\label{subsec:Hc}
%%%%%%%%%%%%%%%%%%%%%%%%%%%%%%%%%%%%%%%%%%%%%%%%%%%%
In this section, we calculate the upper critical 
fields for $H\perp \hat{z}$.
In this case, the parameter $\Lambda$ which characterizes the anisotropy in the
$R_xR_y$-plane is $\Lambda=1$.
The other parameter $\vecc{Q}$ which should be optimized is
$\vecc{Q}=0$ because, for $\vecc{Q}$ to be finite, the interband
pairing on the split Fermi surface is required. However, such
a pairing is energetically
unfavorable.

\begin{figure}[htb]%
\begin{center}
\includegraphics[width=.3\linewidth,height=.225\linewidth]{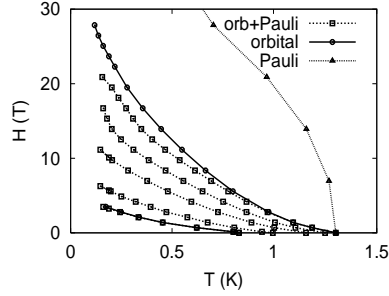}
\end{center}
\caption{%
$H_{c2}(T)$ at $g^2\chi_0/t_1=10$ for several $\theta$. 
The dotted curve with triangles is Pauli limiting field $H_{\rm P}$ 
for $\theta/t_1=0.002$ and
the solid curves with circles are orbital 
limiting fields $H_{\rm orb}$ for 
$\theta/t_1=0.002, 0.03$.
The dotted curves with squares 
are $H_{c2}$ curves including both the Pauli and
the orbital depairing effects for
$\theta/t_1=0.002,0.005,0.01,0.02,0.03$
from the up to the bottom. The $H_{\rm orb}$ curve for $\theta/t_1=0.03$
coincides with the $H_{c2}$ curve with squares.
}
\label{fig:h_c}
\end{figure}

To study $H_{c2}$, we fix the strength of the coupling 
constant as $g^2\chi_0=10t_1$. 
In this calculation, the admixture of the singlet and the triplet
components of the gap functions is fully taken into account,
which is neglected in the previous paper.~\cite{pap:TadaPRL}
The results are almost unchanged from the previous ones
even if we include the effect of
the admixture.
In Fig.\ref{fig:h_c}, $H_{c2}\parallel \hat{z}$ curves as functions of
$T$ for several $\theta$ are shown.
%%%%%%%%%%%%%%%%%%%%%%%%%%%%%%%
%%%%%%%%%%%%%%%%%%%%%%%%%%%
The Pauli limiting field $H_{\rm P}$ is large
because the Rashba SO interaction is strong, 
$\alpha =0.5t_1 > T_c(H=0)\sim 0.01t_1$.
In such a case, the quasiparticles are easily paired on the 
same band under the applied field $\mu_BH\ll \alpha$.
This holds generally and 
does not depend on the symmetry and the dominant parity
of the gap functions for Rashba superconductors.~\cite{pap:Frigeri,
pap:FujimotoJPSJ1,pap:FujimotoJPSJ2}
The upper critical field is, therefore, mainly determined by the orbital 
limiting field $H_{\rm orb}$.
However, as seen in Fig.\ref{fig:h_c}, the orbital limiting field
$H_{\rm orb}$ is different from $H_{c2}$ calculated with both the Pauli 
and the orbital depairing effect being taken into account,
especially for large $H$.
It would be natural to
think that this difference is a numerical artifact due to our
choice of parameters.
The magnitude of $\alpha$ used in the above calculations
is not sufficiently large for high $H$ regions.
If one uses the large value of $\alpha/T_c^{(0)}$, 
where $T_c^{(0)}=t_1\exp [-1/(\rho_0g^2\chi_0)]$ with
the density of states at the Fermi level $\rho_0$,
this difference may disappear.
In fact,
in the experimental data of $H_{c2}$ both in CeRhSi$_3$ and CeIrSi$_3$,
no clear Pauli depairing effect can be seen,
which implies that the Zeeman effect is effectively negligible in the 
compounds.
However, to carry out the numerical calculations of $H_{c2}$ for
the larger $\alpha/T_c^{(0)}$, we need the large size of the
$k$-mesh and a large number of Matsubara frequencies.
We have also calculated $H_{c2}\parallel \hat{z}$ for $g^2\chi_0/t_1=15$.
Although,
in this case, the Zeeman effect much affects $H_{c2}$ compared with 
the $g^2\chi_0/t_1=10$ case, the qualitative behavior of $H_{\rm orb}$
is unchanged.

The calculated $H_{c2}$ show (i) strong $\theta$ dependence
and (ii) upward curvatures, and (iii) they reach $\sim 30$(T).
All these characteristic behaviors well explain the experimental
observations in CeRhSi$_3$ and CeIrSi$_3$ discussed in 
Sec.\ref{sec:intro}.~\cite{pap:Kimura_Hc2,pap:Settai}
The physical reason for these characteristic behaviors in
$H_{c2}\parallel \hat{z}$ is quite simple.
Because, for $H\parallel \hat{z}$, $H_{c2}$ is determined mainly by
the orbital depairing effect and the orbital limiting field $H_{\rm orb}$
can be strongly enhanced by the spin fluctuations near the QCP.
In the quantum critical regime, 
as $T$ is decreased below $T_c(H=0)$, the pairing interaction 
$V\propto \xi^2(T)$
is increased in magnitude while the inelastic scattering between
electrons is suppressed and the quasiparticle damping is decreased,
$\gamma(T) =-{\rm Im}\Sigma_0^R(T) \rightarrow 0$.
This contrasting behaviors of the pairing interaction and
the depairing effect lead to the large enhancement
in $H_{c2}$ for $T\rightarrow 0$ near the QCP.
On the other hand, as discussed in the next section,
the Pauli limiting field $H_{\rm P}$ is not so strongly enhanced 
by the spin fluctuations at low temperatures.
This is a key to resolve the apparent contradiction that
although there are many heavy fermion compounds which are considered to be
located near magnetic QCPs, they do not show such a huge $H_{c2}$
as in CeRhSi$_3$ and CeIrSi$_3$.
In usual centrosymmetric heavy fermion superconductors,
$H_{c2}$ is considered to be mainly determined by the Pauli depairing
effect. Therefore, even if the system is close to the QCP,
$H_{c2}$ is not anomalously enhanced.

\begin{figure}[htb]%
\begin{center}
\includegraphics[width=.3\linewidth,height=.225\linewidth]{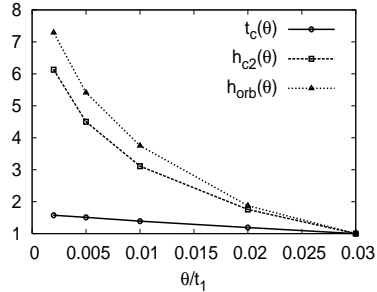}
\end{center}
\caption{%
$t_c$ and $h_{c2}$ as functions of $\theta$.
The dotted curve with triangles 
is calculated from $H_{\rm orb}$,
and the dotted curve with squares includes both
the Pauli depairing and the orbital effects.
The definitions of $t_c(\theta)$ and $h_{c2}(\theta)$ are
given in the text.
}
\label{fig:th-ht}
\end{figure}

The pressure($\theta$) dependence of $H_{c2}(T\rightarrow 0)$
shows a remarkable feature as a result of above mentioned mechanism.
We define normalized $T_c$ and $H_{c2}$ as functions of $\theta$,
$t_c(\theta)\equiv T_c(H=0,\theta)/T_c(H=0,\theta=\theta_{\rm M})$
and $h_{c2}(\theta)=H_{c2}(T=T_{\rm m},\theta)/H_{c2}(T=T_{\rm m},
\theta=\theta_{\rm M})$ where $\theta_{\rm M}=0.03t_1$ and
$T_{\rm m}=0.002t_1$.
The normalized orbital limiting field $h_{\rm orb}$ is also defined in the
same way.
In Fig.\ref{fig:th-ht}, $t_c,h_{c2}$ and $h_{\rm orb}$ 
are shown for $g^2\chi_0=10t_1$.
For $h_{c2}$, the dotted curve with triangles 
is calculated from $H_{\rm orb}$,
and the dotted curve with squares includes both
the Pauli and the orbital depairing effect.
%%%%%%%%%%%%%%%%%%%%%%%%%%%%%%%
%%%%%%%%%%%%%%%%%%%%%%%%%%%
The $\theta$ dependence of $t_c$ is moderate, while
those of both $h_{c2}$ and $h_{\rm orb}$ are significant.
As explained above,
these behaviors are understood as a result of the strongly
enhanced pairing interaction and the suppression of 
the depairing effect at low temperatures
in the vicinity of the QCP($\theta=0$).
Since, in CeRhSi$_3$ and CeIrSi$_3$, the SO interaction
makes the superconductivity orbital limited,
the huge $H_{c2}$ is a result of the interplay of the
Rashba SO interaction and the electron correlations.
Generally,
such strong enhancement in the pairing interaction and 
the suppression of the quasiparticle damping at low temperatures are
crucial for orbital limited superconductors, because
$H_{\rm orb}$ is largely affected by the electron correlations
compared with $H_{\rm P}$.
Therefore, the enhanced upper critical field can be considered
as a universal property of the orbital limited superconductors
near QCPs.
This would be related to the recent experiments of 
$H_{c2}\parallel a$-axis in UCoGe 
in which the relation between the superconductivity and
the ferromagnetism has been discussed.~\cite{pap:UCoGe1}
The observed $H_{c2}^a$ is huge $\sim
15$ (T) while $T_c\sim 1$ (K).~\cite{pap:UCoGe2,pap:UCoGe3}
This issue is now under investigation.

%%%%%%%%%%%%%%%%%%%%%%%%%%%%%%%%%%%%%%%%%%%%%%%%%%%%%
\subsection{$H\perp c$-axis case}
\label{subsec:Hab}
%%%%%%%%%%%%%%%%%%%%%%%%%%%%%%%%%%%%%%%%%%%%%%%%%%%%%%
We also study $H_{c2}$ for the case of $H\perp \hat{z}$
within the same framework.
Since in this case, the Fermi surface is distorted asymmetrically
by the in-plane filed through the Rashba SO interaction,
the Pauli depairing effect is significant, which implies
that the higher Landau levels become important.
Furthermore, the optimization parameter $\vecc{Q}$
and $\Lambda$ are nontrivial for $H\perp \hat{z}$.
First, at a fixed $H$, 
we optimize $\Lambda$ which corresponds to the anisotropy 
in the quasiparticle velocity of the two 
directions perpendicular to the applied field, or
the anisotropy in the superconducting coherence length.
Since $\Lambda$ is characterized by the shape of the Fermi surface,
the field dependence of $\Lambda$ is very weak.
We fix the optimized $\Lambda$, and then,
optimize $\vecc{Q}$ to have the maximum
$H_{c2}$ for given temperatures.
The optimal $\Lambda$ is $\Lambda \simeq 2.3$.

\begin{figure}[htb]%
\begin{center}
\includegraphics[width=.3\linewidth,height=.225\linewidth]{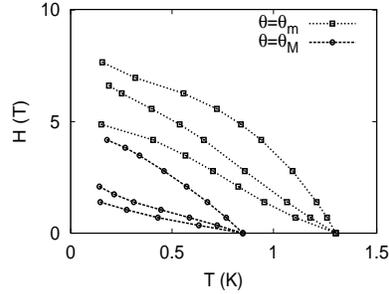}
\end{center}
\caption{%
$H_{c2}\perp \hat{z}$ at $g^2\chi_0/t_1=10$ for $\theta=
\theta_{\rm m}$ (square symbols) and 
$\theta=\theta_{\rm M}$ (circle symbols).
For each $\theta$, three curves correspond to
$N=0,1,2$ Landau levels from the top to the bottom.
}
\label{fig:h_ab_10}
\end{figure}

In Fig.\ref{fig:h_ab_10}, 
we show $H_{c2}$ %for the same parameters as in the previous section 
at $g^2\chi_0=10t_1$
for two values of $\theta$, $\theta_{\rm M}/t_1=0.03$
and $\theta_{\rm m}/t_1=0.002$.
Each $H_{c2}$ curve is calculated with a single Landau function
for $N=0,1,2$,
respectively.
True $H_{c2}$ curve should be calculated by a superposition
of the Landau functions.
We have computed a $H_{c2}$ curve
by using a superposition of $N=0$ and $N=1$ Landau functions,
and found that it almost coincides with the $H_{c2}$ curve calculated
by $N=0$ Landau function only.
%%%%%%%%%%%%%%%%%%%%%%%%%%%%%%%
%%%%%%%%%%%%%%%%%%%%%%%%%%%
Therefore,
$H_{c2}$ is mainly determined by the $N=0$ Landau level, 
and the shapes of $N=0$
$H_{c2}$ curves for $\theta=\theta_{\rm m}$ and $\theta_{\rm M}$
are similar.
This pressure insensitivity is due to the
weak dependence of the Pauli limiting field $H_{\rm P}$
on the electron correlations compared with $H_{\rm orb}$.
The ratio of the calculated value of $H_{c2}(T\rightarrow 0)\perp
\hat{z}$ to that of $H_{c2}(T\rightarrow 0)\parallel \hat{z}$ in the
previous section is $H_{c2}^{\perp}/H_{c2}^{\parallel}\sim 1/3$
for $\theta=\theta_{\rm m}$.
These behaviors in $H_{c2}$ are
consistent with the experiments.~\cite{pap:Kimura_Hc2,pap:Settai}

We turn to the discussion of the modulation vector $\vecc{Q}$.
Under the field $\mu_BH\ll \alpha$, the dispersion is changed as
$\varepsilon_{\tau}(k+q;\vecc{H})\simeq 
\varepsilon_{\tau}(k)+
\vecc{v}_{\tau}(k)\cdot \vecc{q}+
\tau \mu_B\hat{\vecc{\mathcal L}}(k)\cdot \vecc{H}$.
In this situation, the momentum pair $(k_{F\tau}+Q_{\tau},-k_{F\tau})$ is
energetically degenerate on the one band,
where $\vecc{k}_{F\tau}$ is the Fermi momentum for $\tau$-band and
$\vecc{Q}_{\tau}$ satisfies
$\vecc{v}_{\tau}\cdot \vecc{Q}_{\tau}=-\tau
\mu_B \hat{\vecc{\mathcal L}}\cdot \vecc{H}$.
Note that the center of mass momenta of the pairs on each band 
satisfy $\vecc{Q}_+\simeq -\vecc{Q}_-$, and the electrons on each band
favor the center of mass momenta with opposite directions.
Therefore, it is expected that for sufficiently strong $H$,
each band favors each $\vecc{Q}$ and the resulting superconducting
state would be the Fulde-Ferrell-Ovchinnikov-Larkin
(FFLO) state~\cite{pap:FF,pap:LO} with $\Delta \sim 
\eta_{+}\exp[i\vecc{Q}_+\cdot \vecc{R}]
+\eta_{-}\exp[i\vecc{Q}_-\cdot \vecc{R}]$.
For small $H$, however, the helical vortex state with
$\Delta \sim \eta \exp[i\vecc{Q}\cdot \vecc{R}]$ is considered
to be stabilized in general noncentrosymmetric superconductors.
So, the situation is different between the case of small $H$ and 
that of large $H$.
We discuss the $H$-dependence of the 
modulation of the gap function qualitatively.
When the SO split inter-band pairing 
which is small for $\alpha \gg T_c$ is neglected,
the Eliashberg equation for the diagonal element of the
gap $\Delta_{\tau}$ is of the form, 
%%%%%%%%%%%%%%%%%%%%%%%%%%%%
\begin{eqnarray}
\Delta_{\tau}(\alpha,H,Q)&\simeq &
\rho_{+}(\alpha,H){\mathcal F}_{\tau +}(\alpha,H,Q,\Delta_+)
+\rho_{-}(\alpha,H){\mathcal F}_{\tau -}(\alpha,H,Q,\Delta_-)\nonumber \\
&=&\frac{1}{2}(\rho_++\rho_-)({\mathcal F}_{\tau +}+{\mathcal F}_{\tau -})
+\frac{1}{2}(\rho_+-\rho_-)({\mathcal F}_{\tau +}-{\mathcal F}_{\tau -}),
\label{eq:gapeq}
\end{eqnarray}
%%%%%%%%%%%%%%%%%%%%%%%%%%%%%%%
where $\rho_{\tau}$ is the density of states at the Fermi level
for the $\tau$-band
and ${\mathcal F}_{\tau \pm}$ is a function depending on 
$(\alpha,H,Q,\Delta_{\tau})$.
The first term in Eq.(\ref{eq:gapeq}) is proportional to just the sum
$(\rho_{+}+\rho_{-})$
and therefore, the electrons on each band contribute independently.
In contrast, in the second term, the difference between the two bands
plays important roles.
The term is related to the magnetoelectric effect in the
superconducting state due to the anisotropic SO interaction,
which depends on the difference in the densities of
states of the two bands.
This effect is incorporated into the Ginzburg-Landau free energy as
$f_{\rm me}\propto H_{\mu}{\mathcal K}_{\mu \nu}
[\psi^{\ast}D_{\nu}\psi +\psi (D_{\nu}\psi)^{\ast}]$ where 
$D_{\mu}=\partial_{\mu}-i2eA_{\mu}$ and ${\mathcal K}_{\mu  \nu}$
is the coefficient of the magnetoelectric 
effect.~\cite{pap:Edelstein95,pap:Yip,pap:FujimotoPRB,
pap:FujimotoJPSJ1,pap:FujimotoJPSJ2,pap:Kaur,pap:Mineev}
In general noncentrosymmetric superconductors, $f_{\rm me}$ leads to
a spatially modulated gap function 
with the modulation vector $Q_{\nu}^{\rm me}
\propto H_{\mu}{\mathcal K}_{\mu \nu}$.
In this state, the Cooper pair is formed by the states
with $\vecc{k}_{F\tau}+\vecc{Q}^{\rm me}$ 
and $-\vecc{k}_{F\tau}$ momenta, not the 
$(\vecc{k}_{F\tau}+\vecc{Q}_{\tau},-\vecc{k}_{F\tau})$ momenta.
This effect arises even under very weak $H$.
However, it is pointed out that 
in 3D Rashba superconductors in which only ${\mathcal K}_{xy}=
-{\mathcal K}_{yx}$ are nonzero, the phase $\exp[i\vecc{Q}^{\rm me}
\cdot \vecc{R}]$ is
absorbed into the Landau function as a spatial shift
$\phi(\vecc{R})\rightarrow \phi(\vecc{R}-\vecc{R}_0)$ with 
a $H$-dependent vector $\vecc{R}_0$.~\cite{pap:Matsunaga,pap:Hiasa,
pap:Mineev}
Therefore, $\vecc{Q}^{\rm me}$ does not appear in physical 
observables like $H_{c2}$,
although ${\mathcal K}_{xy}$ itself is nonzero.
On the other hand, under a high field, the first term in Eq.(\ref{eq:gapeq})
plays important roles for the optimization of $\vecc{Q}$.
This situation is similar to that of the
FFLO state in usual centrosymmetric superconductors.
Since, as noted above, the first term of Eq.(\ref{eq:gapeq}) is 
a sum of the independent contributions
from two bands, it merely favors $\vecc{Q}_+$ or $\vecc{Q}_-$.
Therefore, in the high field region, the candidate for the
modulation vector is $\vecc{Q}_+$ and $\vecc{Q}_-$.
If $H$ is applied from zero to some large value for general 
noncentrosymmetric superconductors,
we would see a continuous change of $\vecc{Q}$, 
from $\vecc{Q}^{\rm me}$ to $\vecc{Q}_{\pm}$.
The threshold value $H=H^{\ast}$ at which $\vecc{Q}$ changes
from $\vecc{Q}^{\rm me}$ to $\vecc{Q}_{\pm}$ depends on the
details of the system.
It is pointed out that $H^{\ast}$ becomes large as the
orbital depairing effect increases.

\begin{figure}[htb]%
\begin{center}
\includegraphics[width=.3\linewidth,height=.225\linewidth]{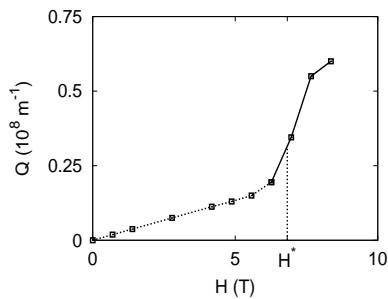}
\end{center}
\caption{%
The modulation $Q$ along the $H_{c2}$ curve for $g^2\chi_0/t_1
=10$ and $\theta=\theta_{\rm m}$.
In the low field region, the optimization of $Q$ would be due to
a numerical artifact and $Q$ is shown by the dotted curve
in this region.
$H^{\ast}$ is the threshold value.
}
\label{fig:h_q_g10}
\end{figure}

In our study, $\vecc{Q}$ is determined so that $H_{c2}$ 
becomes maximum for a given parameter.
We find that
the optimized $\vecc{Q}$ vector is parallel to $a$-axis,
$\vecc{Q}=(-Q,0,0)$ for $\vecc{H}=(0,H,0)$.
In Fig.\ref{fig:h_q_g10}, the optimized $Q$ along 
the $H_{c2}$ curve for $\theta=\theta_{\rm m}$ is shown.
%%%%%%%%%%%%%%%%%%%%%%%%%%%%%%%
%%%%%%%%%%%%%%%%%%%%%%%%%%%
We have two regions; 
the region where $Q$ is 
quite small and the other with large $Q$.
For the small $Q$ region,
although we have a systematic change of $H_{c2}$ with respect to $Q$
and can optimize it,
this dependence of $H_{c2}$ on $Q$ would be a numerical artifact
because the change in $H_{c2}$ due to nonzero $Q$ is infinitesimally small.
Actually, the small $Q$ region corresponding to the helical vortex state
is spurious in a 3D Rashba superconductor because of the reason
mentioned above.
It is expected that the character of the stable vortex state is
nothing but the character of the conventional vortex state 
with $Q=0$ in the region $0<H<H^{\ast}$.
In contrast,
for large $H>H^{\ast}$, we have a finite $Q$.
This large $Q$ state would not be a direct result of the lack of
the inversion center.
Rather, it is stabilized by the pairing of the momentum 
$(\vecc{k}_{F\tau}+\vecc{Q}_{\tau},-\vecc{k}_{F\tau})$ electrons on each band.
However, the contribution to the gap function from the second term
in Eq.(\ref{eq:gapeq}) is not negligible, resulting in the shift
of the degeneracy between $Q_{+}$ and $Q_{-}$.
Therefore, we expect that, in this high field region,
the FFLO state with the gap function $\Delta \sim
\eta_{+}\exp [i\vecc{Q}_+\cdot \vecc{R}]
+\eta_{-}\exp [i\vecc{Q}_-\cdot \vecc{R}]$ can be realized.
We have performed the calculations for other parameters, and 
confirmed that the threshold $H=H^{\ast}$ for the two region
depends on the effective mass $m_{\rm eff}$.
As $m_{\rm eff}$ becomes larger, $H^{\ast}$ decreases, and vice versa,
which means that 
the orbital depairing effect plays important roles for the
determination of $H^{\ast}$.
To discuss the stability of such a state, we need to compute the
free energy in the superconducting state, and it is beyond the 
linearized calculation of $H_{c2}$ performed in the present study.

\begin{figure}[htb]%
\begin{tabular}{cc}
\includegraphics[width=.3\linewidth,height=.225\linewidth]{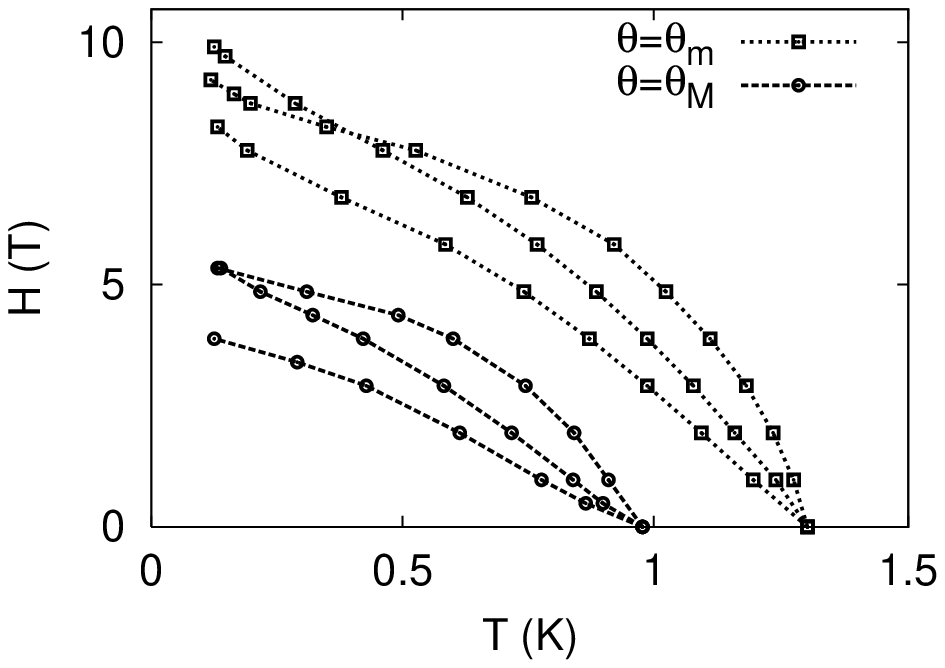}&
\includegraphics[width=.3\linewidth,height=.225\linewidth]{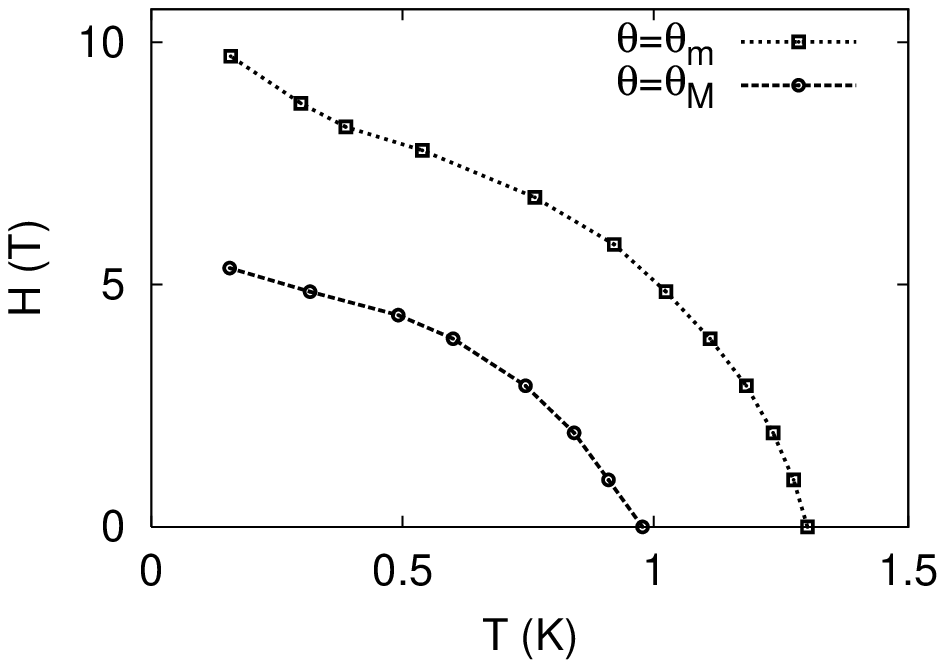}
\end{tabular}
\caption{
$H_{c2}\perp \hat{z}$ at $g^2\chi_0/t_1=15$ for $\theta=
\theta_{\rm m}$ (square symbols) and $\theta=\theta_{\rm M}$
(circle symbols).
Left panel; For each $\theta$, three curves correspond to
$N=0,1,2$ Landau levels from the top to the bottom.
Right panel;
$H_{c2}$ calculated with the use of the superposition of
the $N=0$ and $N=1$ Landau functions.}
\label{fig:h_ab_15}
\end{figure}

As mentioned in Sec.\ref{sec:Eliashberg}, $g^2\chi_0$ cannot be
determined uniquely in our theory, 
and we can change the value of $g^2\chi_0$ within the
range for which the value of $T_c$ is consistent with the experiments.
In the following, we study the case of $g^2\chi_0=15t_1$ which gives
$T_c(H=0,\theta_{\rm m})=0.0199t_1$.
In this case, the value of $t_1$ is $t_1=65.3$(K)
and the effective mass of the cyclotron motion $m_{\rm eff}$ is large
compared with the $t_1=93.8$(K) case.
We show $H_{c2}$ for $\theta=\theta_{\rm m},\theta_{\rm M}$
in Fig.\ref{fig:h_ab_15}.
%%%%%%%%%%%%%%%%%%%%%%%%%%%%%%%%%%%
%%%%%%%%%%%%%%%%%%%%%%%%%%%%%%%%%%%%
The left panel shows $H_{c2}$ curves calculated with
the single Landau functions for $N=0,1,2$, and
the right panel shows $H_{c2}$ curves calculated with
the superpositions of the $N=0$ and the $N=1$ Landau functions.
For $\theta=\theta_{\rm m}$, the $H_{c2}$ curve calcuated with the $N=1$
Landau level is larger than that with the $N=0$ Landau level
at a low temperature region.
In such a region, higher Landau levels become important, and
the gap function can have 
the nodal structure in real space due to
the nodes of the higher Landau functions.~\cite{pap:Matsunaga,pap:Hiasa}
The $H_{c2}$ curves calculated with the use of the superpositions of
the $N=0$ and the $N=1$ Landau levels almost coincide with
the $N=0$ $H_{c2}$ curve for low $H$ and the $N=1$ $H_{c2}$ curve
for high $H$, respectively.
In the case for $\theta =\theta_{\rm m}$, the higher Landau levels becomes
more important than the case for $\theta=\theta_{\rm M}$,
because the orbital depairing effect is largely suppressed
and the electrons are strongly paired near the QCP.

%%%%%%%%%%%%%%%%%%%%%%%%%%%%%%%%%%%%%%%%%%%%%%%%%%%%%%%%
\section{spin-flip scatterings and field dependence of spin fluctuations}
\label{sec:2points}
%%%%%%%%%%%%%%%%%%%%%%%%%%%%%%%%%%%%%%%%%%%%%%%%%%%%%%%%
In the calculation shown in the Sec.\ref{sec:Hc2}, we have neglected
two important effects, 
the spin-flip scattering processes in the pairing interaction
and the field dependence of the spin fluctuations.
Regarding the former, in the noncentrosymmetric systems,
there always exist spin-flip scattering processes
which are not included in Eqs.(\ref{eq:int_1})$\sim$(\ref{eq:int_3}). 
It was pointed out that they can enhance the mixing of the
singlet and the triplet superconductivity,~\cite{pap:Yanase08,pap:Takimoto}
and also, the effective strength of
the Pauli depairing effect depends on the ratio of the 
admixture of the gaps for $H\perp \hat{z}$.~\cite{pap:Hiasa}
Another important point which is neglected in the calculation in 
Sec.\ref{sec:Hc2}
is the field dependence of the susceptibility.
Because the observed $H_{c2}$ is over 20(T) for
$c$-axis in CeRhSi$_3$ and CeIrSi$_3$, one might think that
the spin fluctuations are suppressed by such a strong magnetic
field, although
we have assumed in Eqs.(\ref{eq:int_1})$\sim$(\ref{eq:int_3})
that the spin fluctuations are not strongly 
affected by the magnetic field.
These two points are examined in this section, and it is concluded
that the neglect of them is a legitimate approximation 
and the calculated results in
Sec.\ref{sec:Hc2} are qualitatively unchanged even if we take into
account the two points.

%%%%%%%%%%%%%%%%%%%%%%%%%%%%%%%%%%%%%%%%%%%
\subsection{spin-flip scatterings in pairing interaction}
\label{subsec:flipping}
%%%%%%%%%%%%%%%%%%%%%%%%%%%%%%%%%%%%%%%%%%
In this section, the effects of the spin-flip scattering
processes in the pairing interaction on the superconductivity
are examied.
Through a spin-flip process, such as the scattering process
in which
spin $\uparrow \downarrow$ particles are scattered as spin $\uparrow \uparrow$
particles, the singlet and the triplet pairing
states are mixed directly.
It is pointed out by several authors that
this effect can enhance the admixture of 
the parity even and odd pairing.~\cite{pap:Yanase08,pap:Takimoto}
It is also discussed that, for in-plane fields, 
the effective strength of the Pauli depairing effect depends
on the ratio of the triplet gap function to the 
singlet gap function.~\cite{pap:Hiasa}
In the following, we show that, in CeRhSi$_3$
and CeIrSi$_3$, the admixture of the gap functions
is not so strong even if we include the spin-flip
scattering processes in the pairing interaction.

To investigate the effect of the spin-flip, 
we use the single band Hubbard model
%%%%%%%%%%%%%%%%%%%%%%%%%%%%%%%%%%%%%%%%%%%%%%%%%
\begin{eqnarray}
H=\sum_{k}c_{k}^{\dagger}\varepsilon_0(\vecc{k})c_k
+\alpha \sum_k c_{k}^{\dagger}\vecc{\mathcal L}_0(\vecc{k},\vecc{H})
\cdot \vecc{\sigma}c_k+U\sum_in_{i\uparrow}n_{i\downarrow}.
\label{eq:Hubbard}
\end{eqnarray}
%%%%%%%%%%%%%%%%%%%%%%%%%%%%%%%%%%%%%%%%%%%%%%%%%%%
Here, as in eq.(\ref{eq:action}),
$c_{ks}$ is the annihilation operator of the Kramers
doublet of the heavy electrons which are formed through the hybridization
with the conduction electrons.
The dispersion relation $\varepsilon_0(\vecc{k})$ and 
the Rashba SO interaction
$\vecc{\mathcal L}_0(\vecc{k},\vecc{H})$ are 
defined in Eqs.(\ref{eq:dispersion})
and (\ref{eq:Rashba}).
We fix the parameters as $(t_1,t_2,t_3,t_4,n,\alpha)=
(1.0,0.5,0.3,0.025,0.975,0.2)$ in this 
section and the next section.
The pairing interaction is evaluated by the random phase approximation (RPA)
%%%%%%%%%%%%%%%%%%%%%%%%%%%%%%%%%%%%%%%%%%
\begin{eqnarray}
V_{\alpha \beta, \alpha^{\prime} \beta^{\prime}}(k,k^{\prime})=
U_{\alpha \beta, \alpha^{\prime} \beta^{\prime}}
+[\hat{U}_+\hat{\chi}(k+k^{\prime})\hat{U}_+]
_{\beta^{\prime} \alpha, \beta \alpha^{\prime}}
-[\hat{U}_-\hat{\chi}(k-k^{\prime})\hat{U}_-]
_{\alpha \alpha^{\prime},\beta^{\prime}\beta}
\label{eq:int_RPA}
\end{eqnarray}
%%%%%%%%%%%%%%%%%%%%%%%%%%%%%%%%
where the matrices are defined with the notation
%%%%%%%%%%%%%%%%%%%%%%%%%%%%%%%%%%
\begin{eqnarray}
\hat{M}&=&\left[
\begin{array}{cccc}
M_{\uparrow \uparrow \uparrow \uparrow}
&M_{\uparrow \uparrow \uparrow \downarrow}
&M_{\uparrow \uparrow \downarrow \uparrow}
&M_{\uparrow \uparrow \downarrow \downarrow} \\
M_{\uparrow \downarrow \uparrow \uparrow}
&M_{\uparrow \downarrow \uparrow \downarrow}
&M_{\uparrow \downarrow \downarrow \uparrow}
&M_{\uparrow \downarrow \downarrow \downarrow} \\
M_{\downarrow \uparrow \uparrow \uparrow}
&M_{\downarrow \uparrow \uparrow \downarrow}
&M_{\downarrow \uparrow \downarrow \uparrow}
&M_{\downarrow \uparrow \downarrow \downarrow} \\
M_{\downarrow \downarrow \uparrow \uparrow}
&M_{\downarrow \downarrow \uparrow \downarrow}
&M_{\downarrow \downarrow \downarrow \uparrow}
&M_{\downarrow \downarrow \downarrow \downarrow} \\
\end{array}
\right].
\end{eqnarray}
%%%%%%%%%%%%%%%%%%%%%%%%%%%%%%%
The matrices $\hat{U},\hat{U}_+$ and $\hat{U}_-$ are defined
as
%%%%%%%%%%%%%%%%%%%%%%%%%%%%%%%%%%
\begin{eqnarray}
\hat{U}&=&\left[
\begin{array}{cccc}
0&0&0&0 \\
0&U&-U&0 \\
0&-U&U&0 \\
0&0&0&0 \\
\end{array}
\right],\\
\hat{U}_+&=&\left[
\begin{array}{cccc}
0&0&0&-U \\
0&U&0&0 \\
0&0&U&0 \\
-U&0&0&0 \\
\end{array}
\right],\\
\hat{U}_-&=&\left[
\begin{array}{cccc}
0&0&0&U \\
0&0&-U&0 \\
0&-U&0&0 \\
U&0&0&0 \\
\end{array}
\right].
\end{eqnarray}
%%%%%%%%%%%%%%%%%%%%%%%%%%%%%%%%
The susceptibility $\hat{\chi}(q)$ within RPA is
%%%%%%%%%%%%%%%%%%%%%%%%%%%%%%%%%%
\begin{eqnarray}
\hat{\chi}(q)&=&\hat{\chi}^0(q)[1-\hat{U}_+\hat{\chi}^0(q)]^{-1},\\
\chi^0_{\alpha \beta \alpha^{\prime}\beta^{\prime}}(q)
&=&-\frac{T}{N}\sum_kG^0_{\alpha^{\prime}\alpha}(k)
G^0_{\beta \beta^{\prime}}(k+q).
\end{eqnarray}
%%%%%%%%%%%%%%%%%%%%%%%%%%%%%%%
Equation(\ref{eq:int_RPA}) includes the spin-flip scattering
processes, even for $V_{ssss}$ and $V_{s\bar{s}s\bar{s}}$ as
the virtual scattering processes.
The matrix interation $\hat{V}$
is characterized by the susceptibility 
$\hat{\chi}$, and, 
in the limit of $\alpha \rightarrow 0$, it coincides with
Eqs.(\ref{eq:int_1})$\sim$(\ref{eq:int_3}) if we neglect
the onsite repulsive term $U$ and the charge susceptibility terms.
As shown in Sec.\ref{subsec:susc}, $\hat{\chi}(q)$ has a peak
around $\vecc{q}\sim (\pm 0.5\pi,0,0.5\pi)$ and
$\vecc{q}\sim (0,\pm 0.5\pi,0.5\pi)$ which is consistent with the
neutron scattering experiments for CeRhSi$_3$, and
the $q$-dependence of $\hat{\chi}(q)$ is almost the same
as the phenomenological $\chi(q)$ defined by Eq.(\ref{eq:chi}).

\begin{figure}[htb]%
%\begin{center}
%\includegraphics[width=.8\linewidth,height=.8\linewidth]{test.eps}
%\hspace{-0.7cm}
\begin{tabular}{cc}
\includegraphics[width=.25\linewidth,height=.25\linewidth]{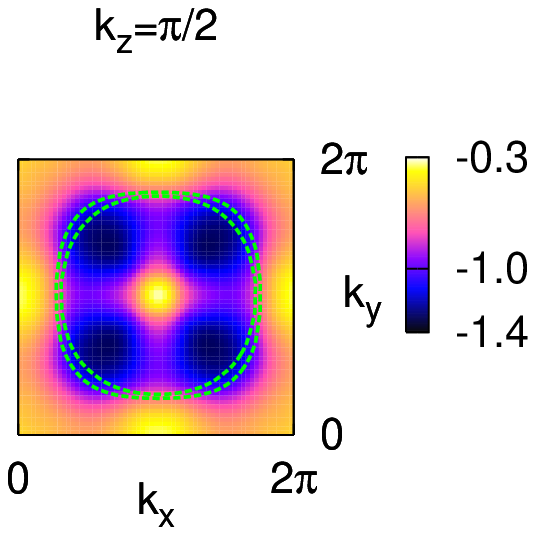}&
\includegraphics[width=.25\linewidth,height=.25\linewidth]{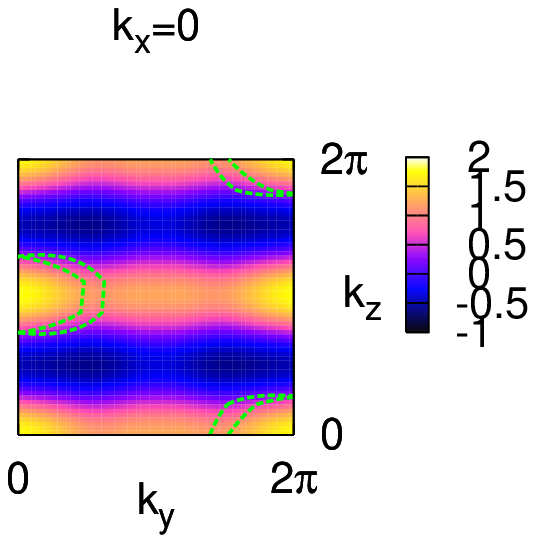}
\end{tabular}
%\end{center}
\caption{%
The singlet gap functions $d_d(\vecc{k})$ in $k_z=\pi/2$ plane (left)
and $k_x=0$ plane (right).
The spin-flip scattering processes are fully taken into account.
The broken curves represent the Fermi surface.
}
\label{fig:d0_full}
\end{figure}

\begin{figure}[htb]%
\begin{tabular}{cc}
\includegraphics[width=.25\linewidth,height=.25\linewidth]{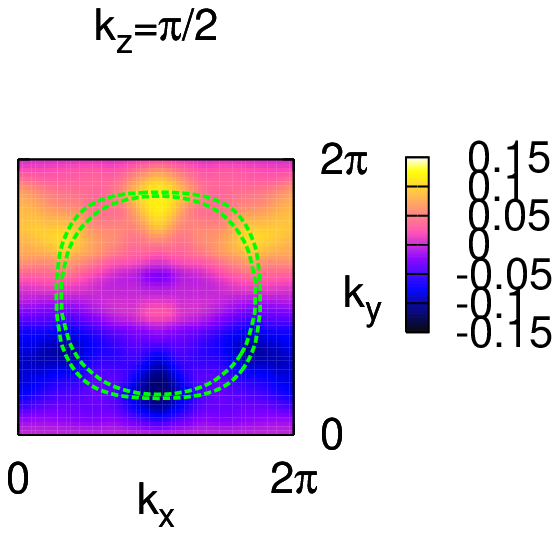}&
\includegraphics[width=.25\linewidth,height=.25\linewidth]{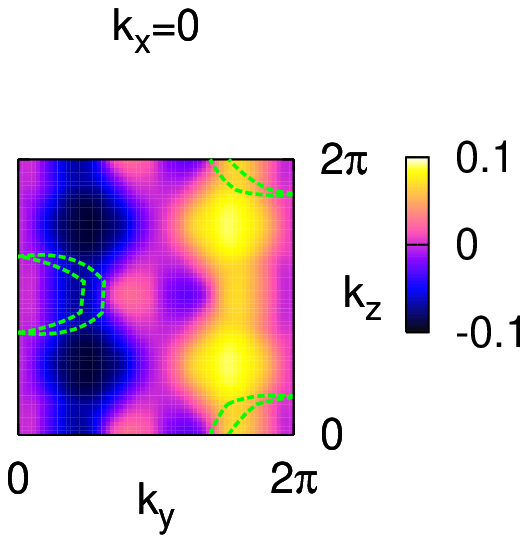}
\end{tabular}
\caption{%
The triplet gap functions $d_1(\vecc{k})$ in $k_z=\pi/2$ plane (left)
and $k_x=0$ plane (right).
The spin-flip scattering processes are fully taken into account.
The broken curves represent the Fermi surface.
}
\label{fig:d1_full}
\end{figure}

\begin{figure}[htb]%
\begin{tabular}{cc}
\includegraphics[width=.25\linewidth,height=.25\linewidth]{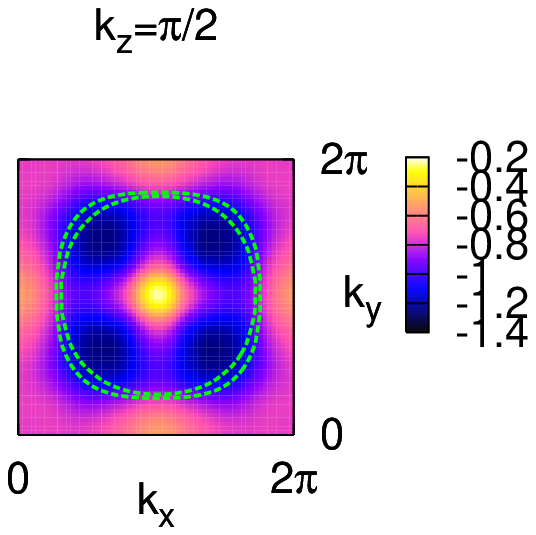}&
\includegraphics[width=.25\linewidth,height=.25\linewidth]{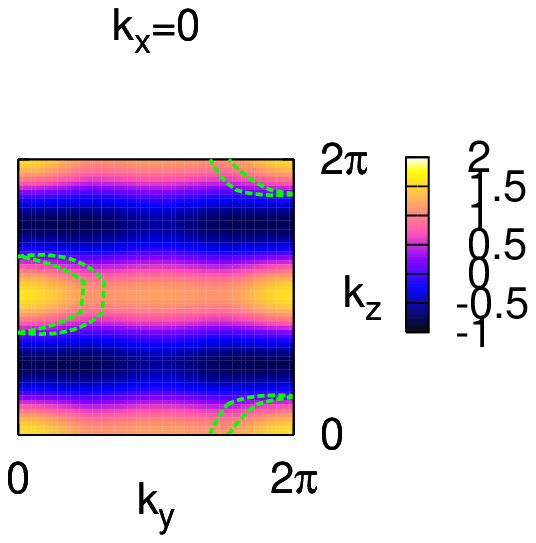}
\end{tabular}
\caption{%
The singlet gap functions $d_0(\vecc{k})$ in $k_z=\pi/2$ plane (left)
and $k_x=0$ plane (right).
The spin-flip scattering processes are not taken into account.
The broken curves represent the Fermi surface.
}
\label{fig:d0_off}
\end{figure}

\begin{figure}[htb]%
\begin{tabular}{cc}
\includegraphics[width=.25\linewidth,height=.25\linewidth]{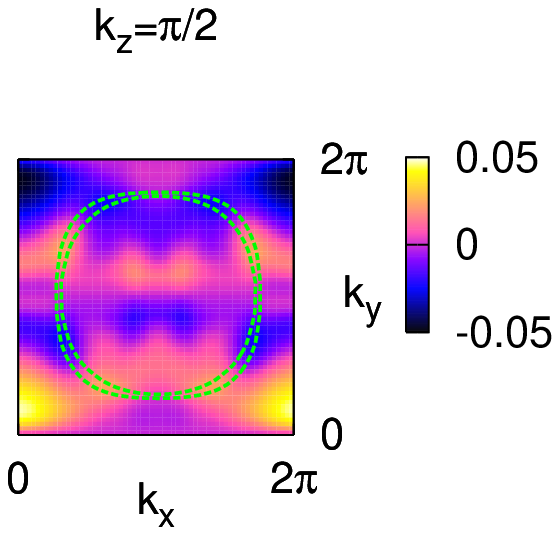}&
\includegraphics[width=.25\linewidth,height=.25\linewidth]{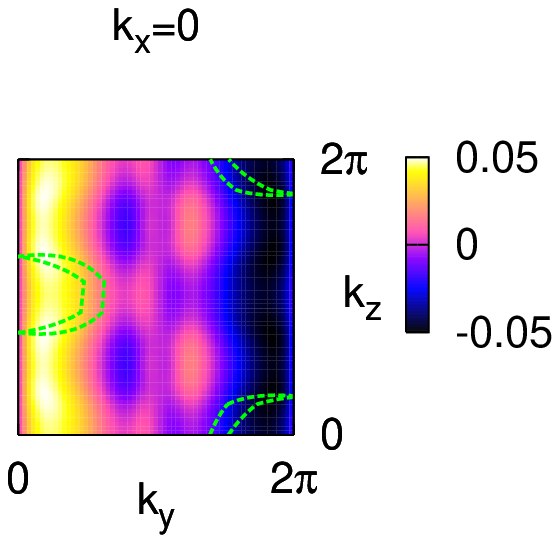}
\end{tabular}
\caption{%
The triplet gap functions $d_1(\vecc{k})$ in $k_z=\pi/2$ plane (left)
and $k_x=0$ plane (right).
The spin-flip scattering processes are not taken into account.
The broken curves represent the Fermi surface.
}
\label{fig:d1_off}
\end{figure}

To discuss the effect of the spin-flip processes on the
admixture of the singlet and the triplet gap functions,
we solve the Eliashberg equation within the weak coupling approximation,
%%%%%%%%%%%%%%%%%%%%%%%%%%%%%%%%%%%%%
\begin{eqnarray}
\Delta_{\alpha \alpha^{\prime}}(\vecc{k})=-\frac{1}{N}\sum_{k^{\prime}}
\tilde{V}_{\alpha \alpha^{\prime}\beta \beta^{\prime}}(\vecc{k},
\vecc{k}^{\prime})g_{\beta \gamma \beta^{\prime}\gamma^{\prime}}
(\vecc{k}^{\prime})
\Delta_{\gamma \gamma^{\prime}}(\vecc{k}^{\prime})
\end{eqnarray}
%%%%%%%%%%%%%%%%%%%%%%%%%%%%%%%%%%%%%%
where $\tilde{V}_{\alpha \alpha^{\prime}\beta \beta^{\prime}}(\vecc{k},
\vecc{k}^{\prime})$ is calculated with the use of 
$\chi_{\alpha \alpha^{\prime}\beta \beta^{\prime}}(i\nu_n=0,\vecc{q})$
%$=V_{\alpha \alpha^{\prime}\beta \beta^{\prime}}(i\omega_n=0,\vecc{k},
%i\omega_m=0,\vecc{k}^{\prime})$
and $g_{\beta \gamma \beta^{\prime}\gamma^{\prime}}
(\vecc{k})=T\sum_{\omega_n}G^0_{\beta \gamma}(k)
G^0_{\beta^{\prime}\gamma^{\prime}}(-k)$.
The pairing interaction $\tilde{V}$ consists of two parts, 
$\tilde{V}_{\rm con}$ corresponding to the spin 
conserving scattering processes,
and $\tilde{V}_{\rm flip}$ including the spin-flip scattering 
processes.
For convenience, we introduce the 4-component $d$-vector of the gap function,
using the identity matrix $\sigma_0$ and the Pauli matrices $\vecc{\sigma}$,
%%%%%%%%%%%%%%%%%%%%%%%%%%%%%%%%%%%%%
\begin{eqnarray}
\Delta (\vecc{k})=\sum_{\mu=0}^3d_{\mu}(\vecc{k})\sigma_{\mu}i\sigma_2,
\end{eqnarray}
%%%%%%%%%%%%%%%%%%%%%%%%%%%%%%%%%%%%
where $d_0$ and $\vecc{d}$ are, respectively, the singlet part
and the triplet part of the gap functions.
We calculate $d_{\mu}(\vecc{k})$ for two cases;
(i) one where all the elements of the interaction matrix
$\tilde{V}_{\alpha \alpha^{\prime}\beta \beta^{\prime}}(\vecc{k},
\vecc{k}^{\prime})$ are fully taken into account, and
(ii) the other where the terms $\tilde{V}_{\rm con}$
including the spin-flip processes are neglected.
We fix the 
parameters as $U=3.5t_1, T=0.02t_1$.
For these parameters, the eigenvalues 
of the Eliashberg equation are $0.95\sim 1.05$.
The gap functions for case (i) are
shown in Fig.\ref{fig:d0_full} for the singlet gap
function $d_0$ and in Fig.\ref{fig:d1_full} for the triplet gap
function $d_1$.
For case (ii),  Fig.\ref{fig:d0_off} and 
Fig.\ref{fig:d1_off} show  $d_0$ and $d_1$,respectively.
%%%%%%%%%%%%%%%%%%%%%%%%%%%%%%%%%%%
%%%%%%%%%%%%%%%%%%%%%%%%%%%%%%%%%%%%
%%%%%%%%%%%%%%%%%%%%%%%%%%%%%%%%%%%
%%%%%%%%%%%%%%%%%%%%%%%%%%%%%%%%%%%%
%%%%%%%%%%%%%%%%%%%%%%%%%%%%%%%%%%%
%%%%%%%%%%%%%%%%%%%%%%%%%%%%%%%%%%%%
%%%%%%%%%%%%%%%%%%%%%%%%%%%%%%%%%%%
%%%%%%%%%%%%%%%%%%%%%%%%%%%%%%%%%%%%
In both cases, the singlet gap function is approximately
$d_0(\vecc{k})\sim \cos (2k_za)$.
The ratio of the triplet gap function to the singlet gap function
defined as $r\equiv |{\rm max}\{d_1(\vecc{k})\}|/|{\rm max}
\{d_0(\vecc{k})\}|$
is about $r\sim 1/10$ for case (i) and $r\sim 1/30$ for case (ii).
Although $r$ is enhanced by the spin-flip processes,
it still remains small in our system.
We have performed similar calculations for various Fermi surfaces,
and found that, generally, the spin-flip processes can enhance $r$.
However, the value of $r$ depends on the details of the system.
In CeRhSi$_3$ and CeIrSi$_3$, we conclude that the admixture of the
gap functions is small.
Therefore, the results in Sec.\ref{sec:Hc2} where we have 
neglected the spin-flip processes in the pairing interaction
are supported.
Finally, we note that the $k$-dependence of the triplet gap function
is largely affected by the spin-flip processes.
If
we write $\tilde{V}=\tilde{V}_{\rm con}+c\tilde{V}_{\rm flip}$ with 
a tuning parameter $c$,
the change in $d_{\mu}$ with respect to $c$ is continuous,
although $d_1$ for case (i) and (ii) look quite different from each other.
This difference in the $k$-dependence in $d_1$ is not
important for the discussion of $H_{c2}$ in Sec.\ref{sec:Hc2}.

%%%%%%%%%%%%%%%%%%%%%%%%%%%%%%%%
\subsection{field dependence of spin fluctuations}
\label{subsec:susc}
%%%%%%%%%%%%%%%%%%%%%%%%%%%%%%%%%%
The field dependence of the spin fluctuations is another effect which
has been neglected in Sec.\ref{sec:Hc2}.
Experimentally observed $H_{c2}$ is so large especially for 
$H_{c2}\parallel \hat{z}$ that one might think that the susceptibility
$\hat{\chi}(q)$ is affected by the applied field and the 
spin fluctuations are weakened.
We show, however, that the effect of the applied field on
$\hat{\chi}(q)$ is strongly suppressed by the Rashba SO interaction.
This is because the Rashba SO coupling tends to fix the direction
of the spins on the Fermi surface depending on $k$-vectors, 
which competes with the Zeeman effect.
As a result, the spin fluctuations in CeRhSi$_3$ and CeIrSi$_3$
are robust against the applied magnetic field up to
the strength of the Rashba SO interaction, $\mu_B H\lesssim \alpha$.

\begin{figure}[htb]%
\begin{tabular}{ccc}
\resizebox{60mm}{50mm}{\includegraphics{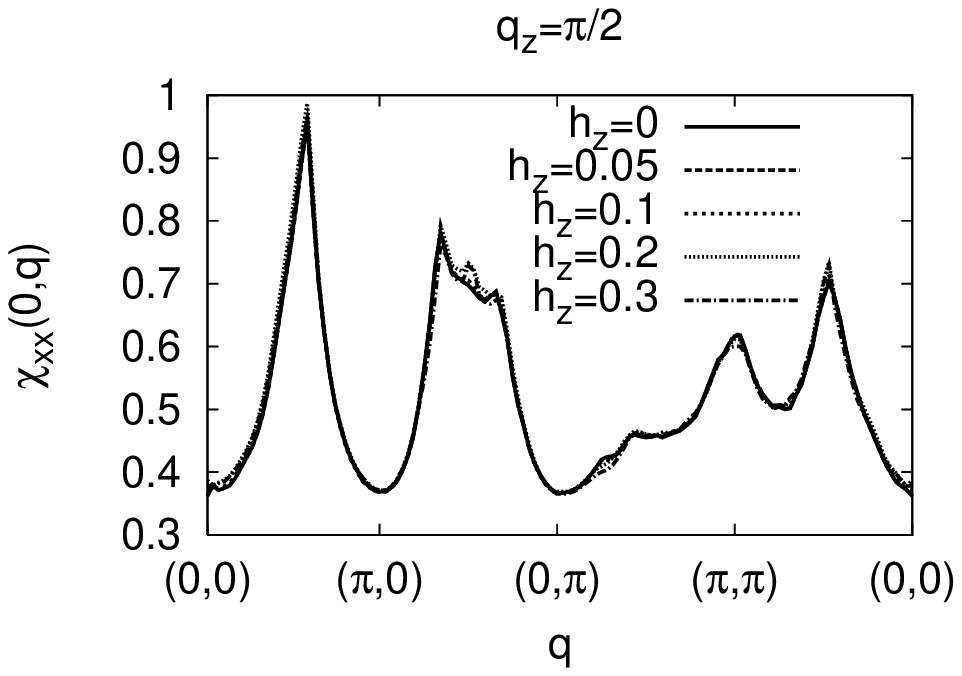}} &
\resizebox{60mm}{50mm}{\includegraphics{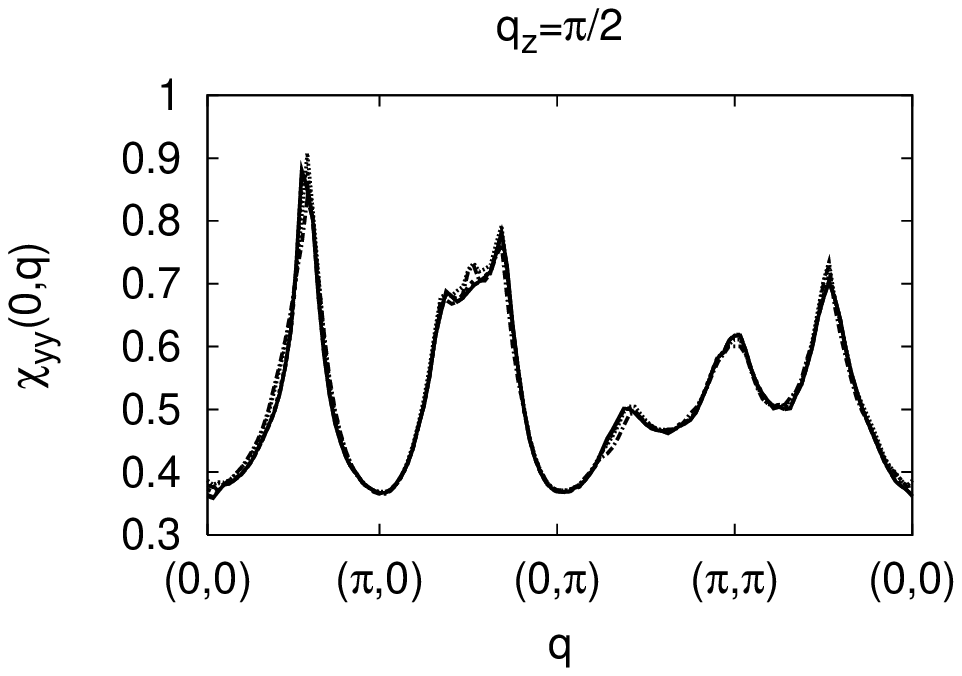}} &
\resizebox{60mm}{50mm}{\includegraphics{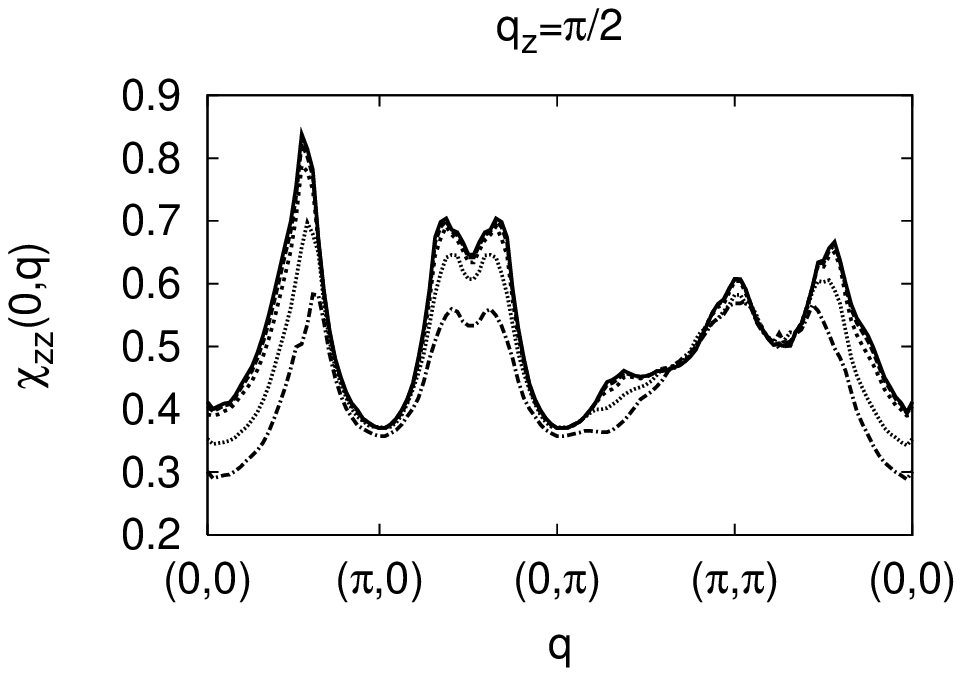}} \\
\end{tabular}
\caption{%
Perpendicular filed $h_z=\mu_B H_z/t_1$ dependence of the
susceptibility $\chi_{\mu \nu}(i\nu_n=0,\vecc{q})$ at $q_z=\pi/2$ for
$U=3.5t_1$ and $T=0.02t_1$.
The strength of the Rashba SO interaction is $\alpha=0.2t_1$.
$\chi_{xx}$(left), $\chi_{yy}$(center), and $\chi_{zz}$(right).
}
\label{fig:chi_hz}
\end{figure}

We compute $\hat{\chi}(q)$ under finite fields 
$\vecc{H}=(0,H_y,0)$ or $(0,0,H_z)$,
%%%%%%%%%%%%%%%%%%%%%%%%%
\begin{align}
\begin{split}
\chi_{\mu \nu}(q)&=\int_0^{1/T}d\tau e^{i\nu_n \tau}
\langle TS^{\mu}_q(\tau)S^{\nu}_{-q}(0)\rangle \\
&=\frac{1}{4}
\sigma^{\mu}_{\alpha \beta}
\chi_{\alpha \beta \beta^{\prime}\alpha^{\prime}}(q)
\sigma^{\nu}_{\alpha^{\prime}\beta^{\prime}}
\end{split}
\end{align}
%%%%%%%%%%%%%%%%%%%%%%%%%%%
where
$\chi_{\alpha \beta \beta^{\prime}\alpha^{\prime}}(q)$ is
evaluated within RPA used in Sec.\ref{subsec:flipping}.
We fix $U$ and $T$, as in the previous section,
$U=3.5t_1$ and $T=0.02t_1$.
In Fig.\ref{fig:chi_hz}, $H$-dependence of $\chi_{xx},\chi_{yy}$ and
$\chi_{zz}$ is shown for $\vecc{H}=(0,0,H_z)$.
%%%%%%%%%%%%%%%%%%%%%%%%%%%%%%%%%%%
%%%%%%%%%%%%%%%%%%%%%%%%%%%%%%%%%%%%
At $H=0$, $\chi_{xx}(0,\vecc{Q}_x)=\chi_{yy}(0,\vecc{Q}_y)
>\chi_{zz}(0,\vecc{Q}_{x,y})$ are satisfied, which is 
consistent with the result of the neutron scattering experiments
in CeRhSi$_3$ that 
the antiferromagnetic moment is in $ab$-plane.~\cite{pap:Aso}
Here, $\vecc{Q}_{x}\sim(\pm 0.5\pi,0,0.5\pi)$ and 
$\vecc{Q}_{y}\sim(0,\pm 0.5\pi,0.5\pi)$. Note that
$\chi_{xx}(0,\vecc{Q}_x)>\chi_{yy}(0,\vecc{Q}_x)$ is 
satisfied because of the spin-flip scattering processes.
However, the anisotropy in the non-interacting 
$\chi^0_{\mu \mu}(q)$ is of the order of $\alpha/\varepsilon_F\ll 1$,
and therefore, the anisotropy in $\chi_{\mu \mu}(q)$ including
the electron correlation effect remains
irrelevant for the discussion of $H_{c2}$ even near the QCP.
As mentioned above,
for $\mu_BH\lesssim \alpha$, $\{ \chi_{\mu \mu}\}$ are almost unchanged.
For $\mu_BH\gtrsim \alpha$, 
only $\chi_{zz}$ is suppressed.
The robustness of $\{ \chi_{\mu \mu}\}$ for $\mu_BH\lesssim \alpha$
is a general feature of the noncentrosymmetric systems,
since the spins for every $k$-point are fixed by the anisotropic SO
interaction in that region.
These calculated results support the legitimacy of
our neglecting the field dependence of the
pairing interaction for the calculation of $H_{c2}$.
Although there is no direct observation of the strength of the
SO interaction, it is expected to be pretty large, 
$\alpha >\mu_BH_{c2}\sim 30\mu_B$(K).
Therefore, in CeRhSi$_3$ and CeIrSi$_3$, the spin fluctuations
remain so strong under applied magnetic fields that $H_{c2}$
is strongly enhanced.

\begin{figure}[htb]%
\begin{tabular}{ccc}
\resizebox{60mm}{50mm}{\includegraphics{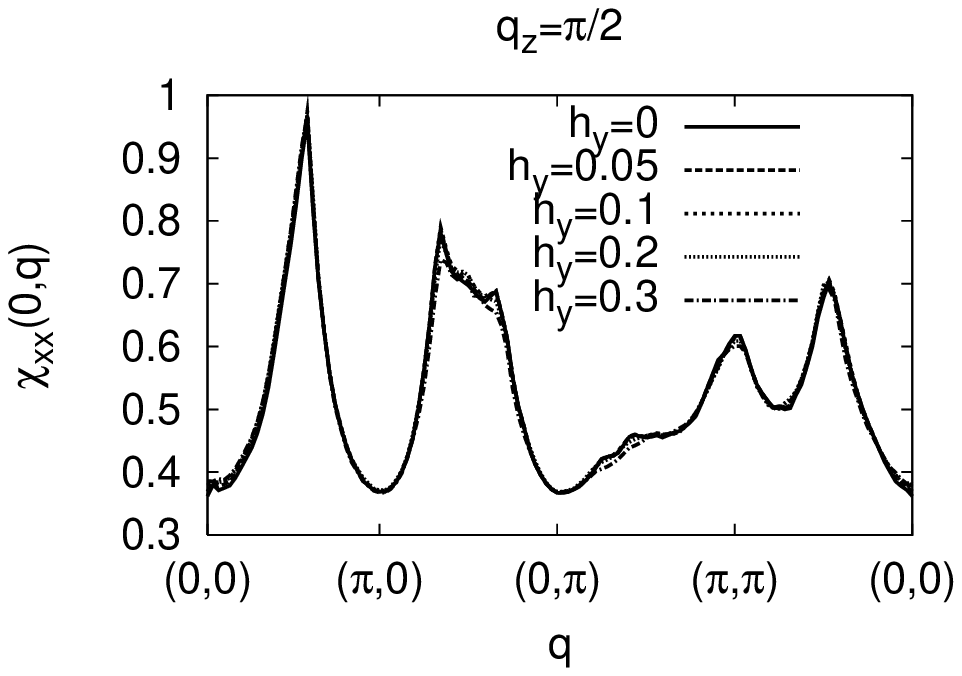}} &
\resizebox{60mm}{50mm}{\includegraphics{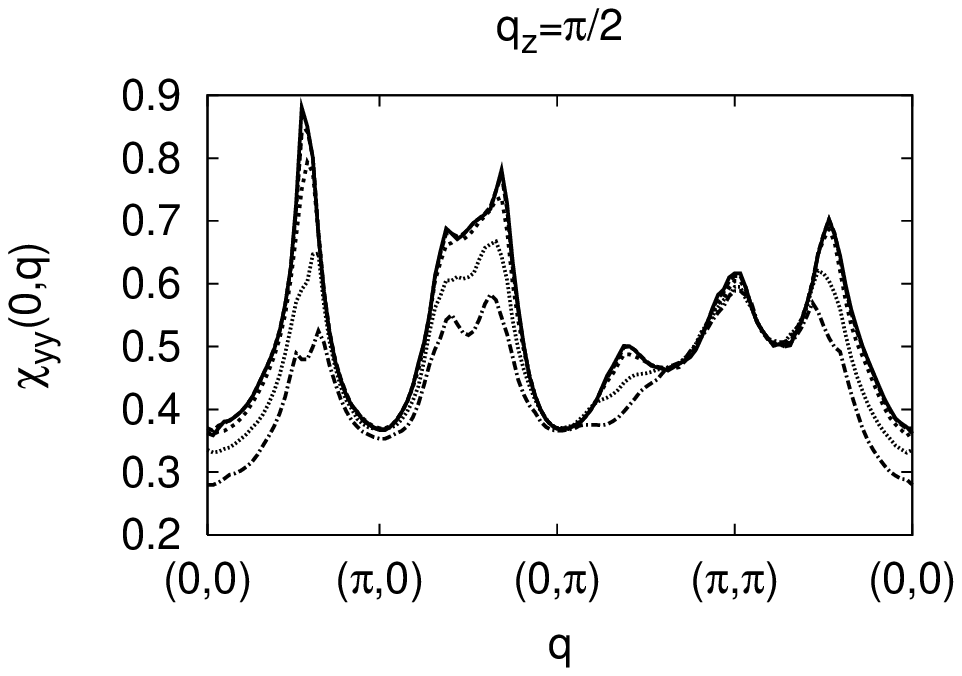}} &
\resizebox{60mm}{50mm}{\includegraphics{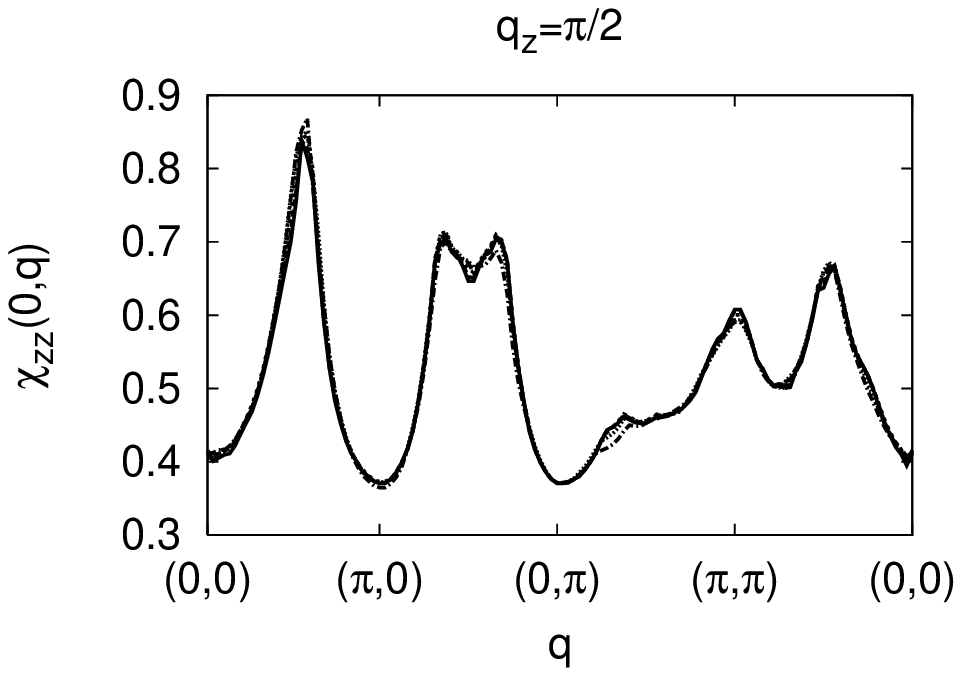}} \\
\end{tabular}
\caption{%
In-plane filed $h_y=\mu_B H_y/t_1$ dependence of the
susceptibility $\chi_{\mu \nu}(0,\vecc{q})$ at $k_z=\pi/2$ for
$U=3.5t_1$ and $T=0.02t_1$.
The strength of the Rashba SO interaction is $\alpha=0.2t_1$.
$\chi_{xx}$(left), $\chi_{yy}$(center), and $\chi_{zz}$(right).
}
\label{fig:chi_hy}
\end{figure}

The same robustness also exists for $\vecc{H}=(0,H_y,0)$ for which
the Fermi surface is distorted anisotropically.
Figure\ref{fig:chi_hy} shows the field dependence of $\chi_{\mu \mu}$.
All of $\chi_{\mu \mu}$ are almost unchanged for $\mu_BH\lesssim \alpha$,
and $\chi_{yy}$ is suppressed for $\mu_BH\gtrsim \alpha$.
%%%%%%%%%%%%%%%%%%%%%%%%%%%%%%%%%%%
%%%%%%%%%%%%%%%%%%%%%%%%%%%%%%%%%%%%
These behaviors are basically the same as those for 
$\vecc{H}\parallel \hat{z}$, and assert the robustness
of the spin fluctuations against the in-plane field.

From the above results, we see that neglecting the field dependence
in the pairing interaction for any field direction is legitimate
provided $\mu_BH_{c2}\lesssim \alpha$. 
Although $H_{c2}$ is huge in CeRhSi$_3$ and CeIrSi$_3$ 
especially for $c$-axis,
the condition $\mu_BH_{c2}\lesssim \alpha$ is expected to be satisfied.
Therefore, the discussion of $H_{c2}$ in Sec.\ref{sec:Hc2}
is not changed even if we consider the field dependence of the
spin fluctuations.

%%%%%%%%%%%%%%%%%%%%%%%%%%%%%%%%%%%%%%%%%%%%%%%%%%%%%%%%%%%%%%%%%%%%%%%%%%%
\section{Summary}
\label{sec:summary}
%%%%%%%%%%%%%%%%%%%%%%%%%%%%%%%%%%%%%%%%%%%%%%%%%%%%%%%%%%%%%%%%%%%%%%%%
We have discussed the normal and the superconducting properties
in noncentrosymmetric heavy fermion superconductors CeRhSi$_3$ and
CeIrSi$_3$.
We have shown that the $T$-linear dependence of the resistivity 
above $T_c$ observed
experimentally is naturally understood within the 3D spin fluctuations
near the AF QCP.

For the superconducting state, we have derived a formula from
the Eliashberg equation in real space.
The formula enables us to treat the Pauli and the orbital depairing
effects on an equal footing.
Furthermore, by using it,
we can calculate $H_{c2}$ for strong coupling superconductors
with general Fermi surfaces.
We have calculated $H_{c2}$ with the formula and 
have well explained the
observed features of $H_{c2}$ in CeRhSi$_3$ and
CeIrSi$_3$.
For $H\parallel \hat{z}$, $H_{\rm P}$ is infinitely large
due to the Rashba SO interaction, and $H_{c2}$ is determined by
$H_{\rm orb}$. As temperature is lowered and the system approaches 
the QCP,
the pairing interaction becomes larger while the quasiparticle
life time becomes longer, which results in the huge 
$H_{\rm orb}\simeq H_{c2}$ with
the strong pressure dependence.
The enhancement of the orbital limiting field near QCPs by this
mechanism would be universal.
We have also discussed the case for $H\perp \hat{z}$.
In this case, the Pauli depairing effect is significant 
because of the asymmetric distortion of the Fermi surface, and
resulting $H_{c2}$ is moderate against the pressure.
The FFLO state can be stabilized for a large $H$ region, 
although such a region is very small.
The features of the calculated $H_{c2}$ for both $H\parallel \hat{z}$
and $H\perp \hat{z}$ are in good agreement
with the experiments.
This consistency supports the scenario that 
the superconductivity in CeRhSi$_3$ and CeIrSi$_3$ is mediated
by the spin fluctuations near the AF QCP.

In the last section, we have checked the legitimacy of our
approximation used for the calculation of $H_{c2}$. 
In CeRhSi$_3$ and CeIrSi$_3$, the admixture of the singlet and the
triplet gap functions are small even if we take into account the
spin-flip scattering processes in the pairing interaction.
In noncentrosymmetric systems, the spin susceptibility is robust
against the applied magnetic fields $\mu_BH\lesssim \alpha$.
For this reason, the spin fluctuations near the AF QCP
in CeRhSi$_3$ and CeIrSi$_3$ remain strong even under a large
magnetic field $\sim 30$ (T).
Therefore,
the above mentioned results for $H_{c2}$ is not
changed if we refine our approximation used in the calculation 
of $H_{c2}$.

%%%%%%%%%%%%%%%%%%%%%%%%%%%%%%%%%%%%%%%%%%%%%%%%%%%%%%%%%%%%
\section*{Acknowledgement}
%%%%%%%%%%%%%%%%%%%%%%%%%%%%%%%%%%%%%%%%%%%%%%%%%%%%%%%%%%%%
We thank N. Kimura, R. Settai and Y. \=Onuki for valuable discussions.
Numerical calculations were partially performed at the 
Yukawa institute. 
This work is partly supported by the Grant-in-Aids for
Scientific Research from MEXT of Japan
(Grant No.18540347, Grant No.19052003, 
Grant No.20029013, Grant No.20102008, Grant No.21102510, and
Grant No.21540359) 
and the Grant-in-Aid for the Global COE Program 
"The Next Generation of Physics, Spun from Universality and Emergence".
Y. Tada is supported by JSPS Research Fellowships for Young
Scientists.

%%%%%%%%%%%%%%%%%%%%%%%%%%%%%%%%%%%%%%%%%%%

\end{document}